\documentclass[letter]{aa} 

\usepackage{graphicx}
\usepackage{txfonts}
\begin{document}

   \title{The wonderful complexity of the Mira AB system}

   \author{S. Ramstedt
          \inst{1} \and S. Mohamed \inst{2} \and W.~H.~T. Vlemmings \inst{3} \and M. Maercker \inst{3} \and R. Montez \inst{4} \and A. Baudry \inst{5} \and E. De Beck \inst{3} \and M.~Lindqvist \inst{3} \and H. Olofsson \inst{3} \and E.~M.~L. Humphreys \inst{6} \and A. Jorissen\inst{7} \and F. Kerschbaum \inst{8} \and A. Mayer \inst{8} \and M. Wittkowski \inst{6}  \and N.~L.~J.~Cox \inst{9} \and E.~Lagadec \inst{10} \and M.~L. Leal-Ferreira \inst{11} \and  C. Paladini \inst{7} \and A. P\'erez-S\'anchez \inst{12} \and S. Sacuto \inst{1}
          }

   \institute{Department of Physics and Astronomy, Uppsala University, Box 516, 75120 Uppsala, Sweden \\
              \email{sofia.ramstedt@physics.uu.se} \and South African Astronomical Observatory, P.O. box 9, 7935 Observatory, South Africa \and Dept. of Earth and Space Sciences, Chalmers University of Technology, Onsala Space Observatory, SE-439 92 Onsala, Sweden \and Department of Physics \& Astronomy, Vanderbilt University, Nashville, TN \and Univ. Bordeaux, LAB, UMR 5804, 33270 Floirac, France \and ESO, Karl-Schwarzschild-Str. 2, 85748 Garching bei M\"unchen,
Germany \and Institut d'Astronomie et d'Astrophysique, Universit\'e Libre de Bruxelles, CP. 226, Boulevard du Triomphe, 1050 Brussels, Belgium \and Dept. of Astrophysics, Univ. of Vienna, T\"urkenschanzstr. 17, 1180, Vienna, Austria \and Instituut voor Sterrenkunde, KU Leuven, Celestijnenlaan, 200D, 3001 Leuven, Belgium \and Laboratoire Lagrange, UMR7293, Univ. Nice Sophia-Antipolis, CNRS, Observatoire de la C\^{o}te d'Azur, 06300 Nice, France \and  Argelander-Institut f\"ur Astronomie, Universit\"at Bonn, Auf dem H\"ugel 71, 53121 Bonn, Germany \and Centro de Radioastronom\'ia y Astrof\'isica, Univ. Nacional Aut\'onoma de 
M\'exico, PO Box 3-72, 58090 Morelia, M\'exico  
             }

   \date{}

 
  \abstract
   {We have mapped the $^{12}$CO(3-2) line emission around the Mira AB system at 0\farcs5~resolution using the Atacama Large Millimeter/submillimeter Array (ALMA). The CO map shows amazing complexity. The circumstellar gas has been shaped by different dynamical actors during the evolution of the system and several morphological components can be identified. The companion is marginally resolved in continuum emission and is currently at 0\farcs487$\pm$0\farcs006 separation. In the main line component, centered on the stellar velocity, spiral arcs around Mira A are found. The spiral appears to be relatively flat and oriented in the orbital plane. An accretion wake behind the companion is clearly visible and the projected arc separation is of order 5\arcsec. In the blue wing of the line emission, offset from the main line, several large ($\sim$5-10\arcsec), opposing arcs are found. We tentatively suggest that this structure is created by the wind of Mira B blowing a bubble in the expanding envelope of Mira A. }


   \maketitle
%

\section{Introduction}
\label{intro}
The Mira AB system is well-studied and nearby \citep[the parallax distance is 92\,pc,][]{vlee07}. It is a regularly pulsating M-type AGB star with a companion detected at 0\farcs6 separation \citep{karoetal97}. The latest calculation of the orbit \citep{prieetal02}, gives an inclination of 112$^{\circ}$, an orbital period of $\sim$500\,yrs, and the current separation is expected to be $\sim$0\farcs5~with a position angle of $\sim$100$^{\circ}$. The mass-loss rate from Mira A is estimated to a few times 10$^{-7}$\,M$_{\odot}$\,yr$^{-1}$ \citep[e.g.,][without taking the asymmetries of the wind into account]{knapetal98,rydescho01} and CO gas emission is detected with local-standard-of-rest (lsr) velocities from 37 to 54\,km\,s$^{-1}$, suggesting a relatively low wind velocity ($\sim$5\,km\,s$^{-1}$).

Several imaging studies have already shown the morphological intricacy of the circumstellar material on different scales. GALEX images revealed how interaction with the ISM has created a 2$^{\circ}$ tail and a bowshock $\sim$3\arcmin~south of the system \citep{martetal07}. They also show a series of knots on a 10\arcmin -scale, later studied in H$\alpha$ \citep{meabetal09}, and referred to as the north and south stream. The streams are tilted relative to the plane of the sky by 69$^{\circ}$ with the north part receding and the south approaching. On a somewhat smaller scale ($\sim$5\arcmin), Herschel/PACS imaged the cold dust \citep{mayeetal11}. The PACS images show the complex circumstellar dust distribution with several large arcs surrounding the system. Previous images of the CO emission show a wide bipolar structure with the north part approaching and the south receding \citep{planetal90,jossetal00,fongetal06}, i.e., perpendicular to the north and south stream. The higher-resolution (2\farcs5) map shows a 20\arcsec~butterfly shape where the molecular emission is believed to have been shaped by a low-velocity atomic outflow \citep{jossetal00}. HST images first resolved the two stars and showed material flowing from Mira A to B \citep{karoetal97}. This overflow was later confirmed by the Chandra X-ray image that shows a weak bridge between the two components \citep{karoetal05}. It has been suggested that the accretion generates a wind from Mira B confirmed by the P-Cygni profile of its strong Mg II $k$ line \citep[e.g.,][]{woodkaro06}. The wind is variable with mass-loss rates ranging from 5$\times$10$^{-13}$\,M$_{\odot}$\,yr$^{-1}$ to 1$\times$10$^{-11}$\,M$_{\odot}$\,yr$^{-1}$, and with terminal wind velocities observed between 250\,km\,s$^{-1}$ and 450\,km\,s$^{-1}$.


We have mapped the CO(3-2) line emission around the system with considerably higher spatial resolution than previous observations ($\sim$0\farcs5). In Sect.~\ref{obs}, we present the observations. In Sect.~\ref{res}, the results are shown and interpreted. In Sect.~\ref{dis}, the morphology is put into the context created by previous observations and possible shaping scenarios are discussed. 

\begin{figure*}[t]
   \centering
   \includegraphics[height=7.6cm]{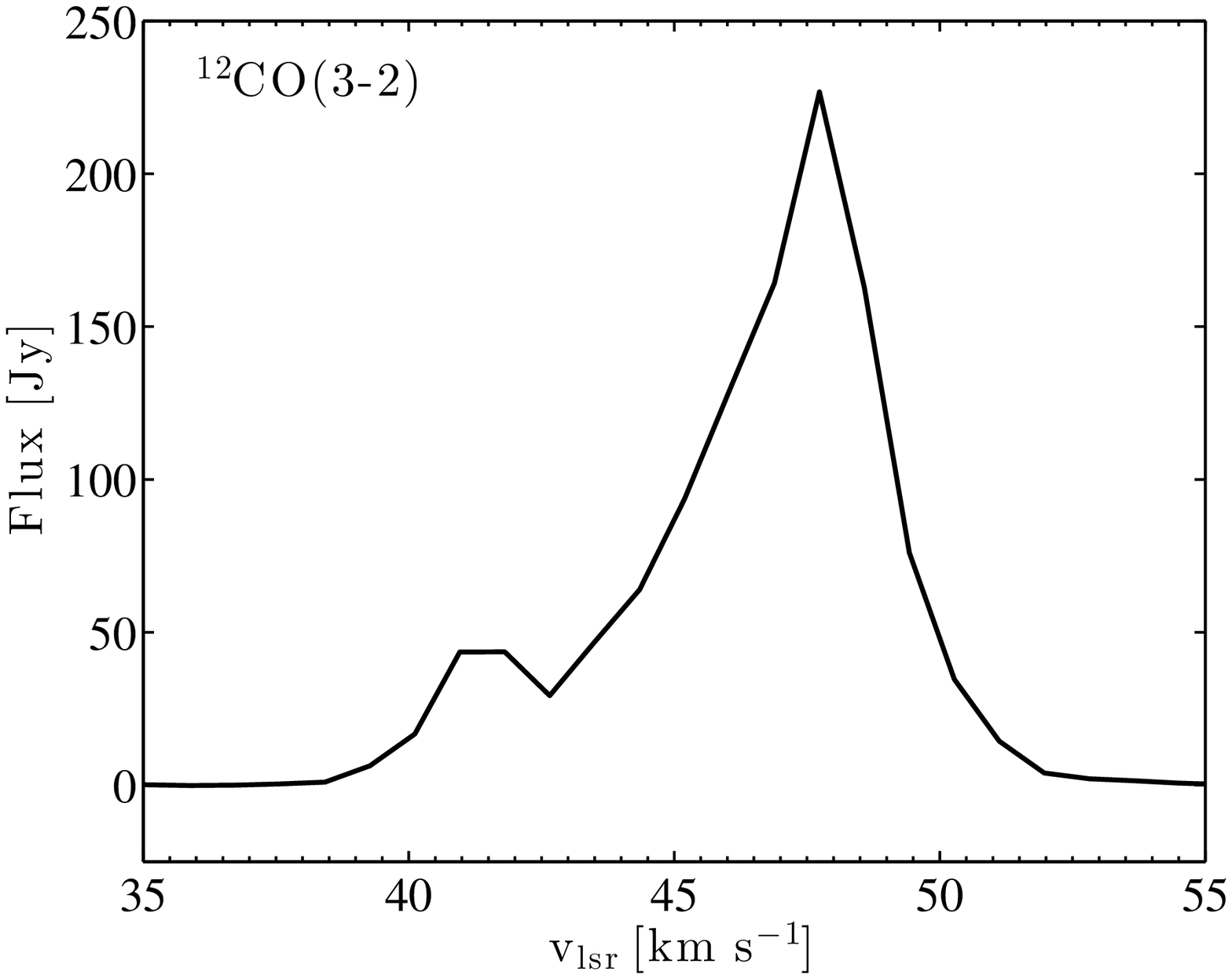}
   \hspace{0.5cm}
   \includegraphics[height=7.4cm]{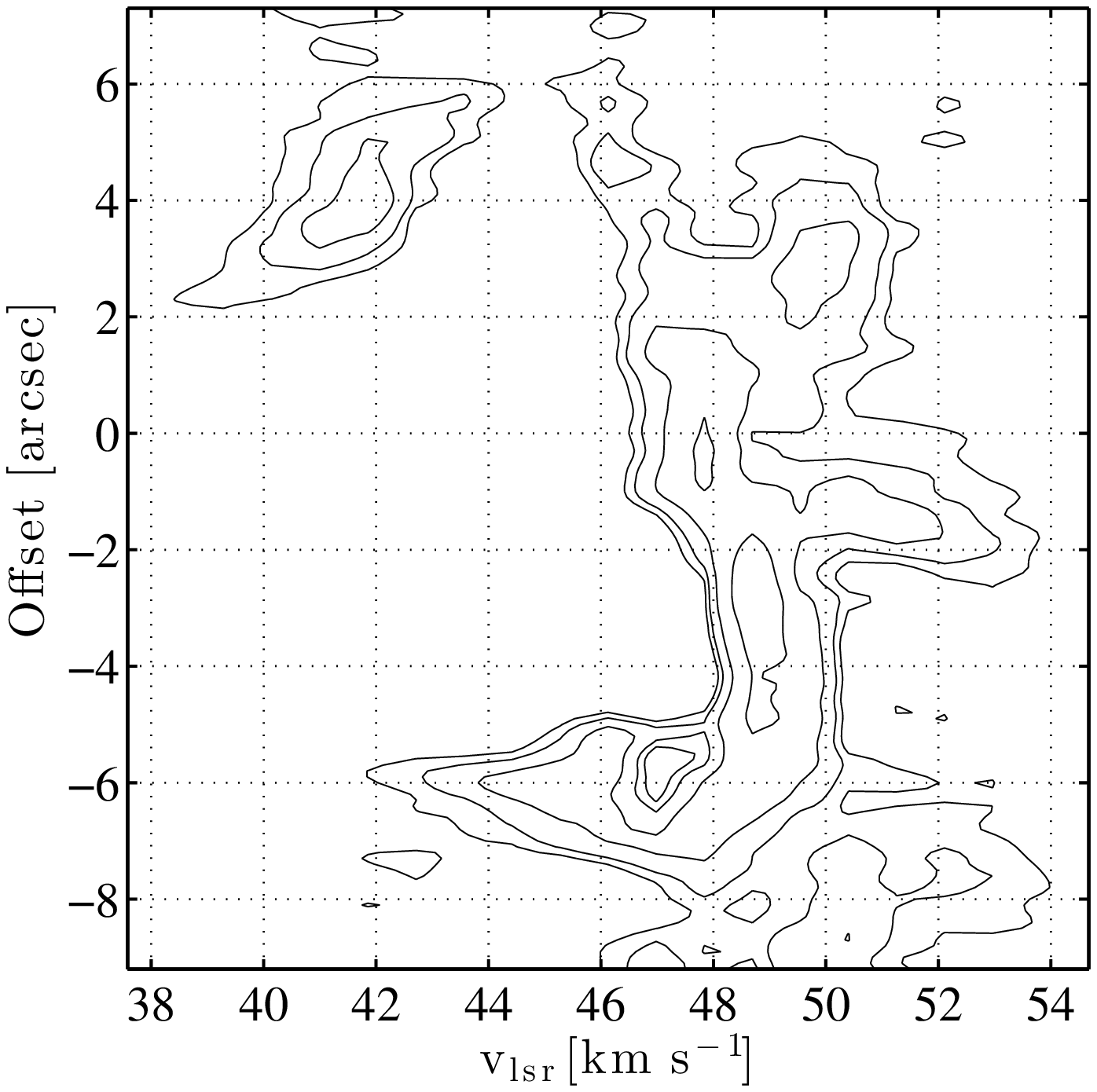} 
   \caption{{\it Left:} The ALMA CO(3-2) line profile generated using an 25\arcsec~beam centered on Mira A. {\it Right:} Position-velocity diagram showing the positional offset relative to the center of the bubble along a position angle of 38$^{\circ}$, as a function of lsr velocity.}
   \label{pv}
\end{figure*}

\section{Observations}
\label{obs}
Mira was observed on 25 February 2014 and 3 May 2014, with 27 and 30 of the main array 12m antennas, respectively. It was also observed with seven Atacama Compact Array (ACA) 7m antennas on 8 October 2013. The main array observations consist of a 10-point mosaic covering an area of $\sim$25\arcsec$\times$25\arcsec. The same area is covered with three mosaic pointings with the ACA. The observations have four spectral windows with a width of $1.875$\,GHz (main array) and $2$\,GHz (ACA) centered on 331, 333, 343 and 345\,GHz. The resulting spectral resolution is $0.488$~MHz ($\sim$0.42\,km\,s$^{-1}$) for the main array and ACA. The total time on source is 6.25~minutes with the main array and 6~minutes with the ACA. The baseline length of the main array ranges from 13 to 450\,m, from which we find a maximum recoverable scale (MRS) of $\sim$6\arcsec. The ACA covers 9 to 45\,m baselines and the MRS is $\sim$9\arcsec.  Calibration on the separate data sets was carried out using CASA, following standard procedures, using the quasar J0217+0144 as complex gain calibrator, quasars J0423-0120 and J006-0623 as bandpass calibrators and Ganymede, Uranus and the quasar J0334-401 as flux calibrators. The data were subsequently combined, weighing the ACA data down by a factor of $0.3$ to account for the lower sensitivity of the 7m antennas. Tests indicate that the exact weight does not affect our results. A continuum image was made using the emission free channels from all spectral windows. As the continuum emission is sufficiently strong, this image was used for self-calibration. After applying the self-calibration, final imaging was done after subtracting the continuum, and averaging over 2 channels to $\sim$0.85\,km\,s$^{-1}$ spectral resolution. The images have a beam full width half maximum (FWHM) of 0\farcs576$\times$0\farcs403 and reach an rms noise of 23 mJy/beam in the emission free channels. 

\begin{table}[htbp]
   \caption{Current positions (ep=J2000.0) and continuum flux density at 338\,GHz, $S_{\nu}$, of Mira A and B determined from the ALMA data.}   
   \begin{tabular}{lccc} 
      \hline \hline
      Source    & R.A. & Dec. & $S_{\nu}$ \\
      \hline
      Mira A & 02:19:20.784 & -02:58:42.818 & 252.3$\pm$0.3\,mJy \\
      Mira B & 02:19:20.816 & -02:58:42.887 & \phantom{1}15.3$\pm$0.3\,mJy \\
      \hline \hline
   \end{tabular}
   \label{contdata}
\end{table}

\begin{figure}[t]
   \centering
   \includegraphics[height=7.4cm]{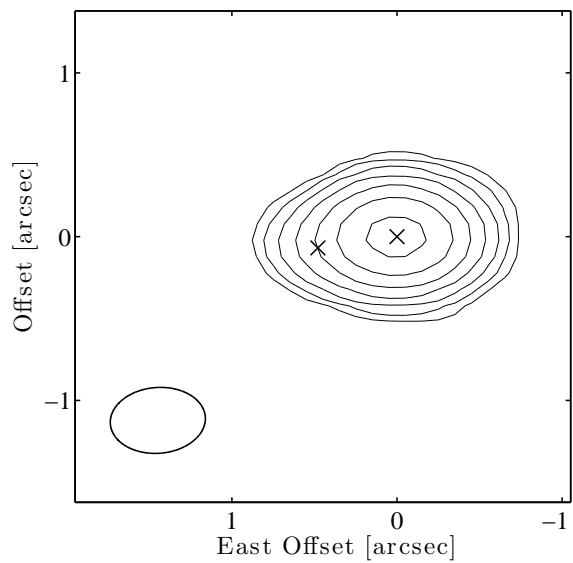}
   \caption{The continuum emission at 338\,GHz with the current positions of Mira A (west) and B (east, marked by crosses). The contours are drawn at 5, 10, 20, 40, 80, 160, and 320 times the rms noise level at 0.6\,mJy/beam. The beam is shown in the lower left corner.}
   \label{cont}
\end{figure}

\section{Results}
\label{res}
\subsection{Continuum emission and current positions}
\label{sep}
The binary pair is marginally resolved when imaged in the continuum channels of the data (Fig.~\ref{cont}). The current positions determined by uv fitting \citep{martetal14} are given in Table~\ref{contdata} together with the measured flux density at 338\,GHz of the two sources. The separation is 0\farcs482$\pm$0\farcs005 in right ascension (R.A.) and 0\farcs069$\pm$0\farcs003 in declination (Dec.). This gives a total separation of 0\farcs487$\pm$0\farcs006. The flux density of Mira B fits quite well with the VLA radio continuum spectral index determination of \citet{mattkaro06}, however that of Mira A is stronger by a factor of a few. In order for the $S_{\nu}$ $\propto$ $\nu^{1.5}$ relation determined by \citet{mattkaro06} for Mira A to hold to submm wavelengths, Mira A radio continuum must have become stronger by a factor $\gtrsim$4 since 2005.

\subsection{Circumstellar morphology}
\label{morph}
\begin{figure*}[htbp]
   \centering
   \includegraphics[height=7.5cm]{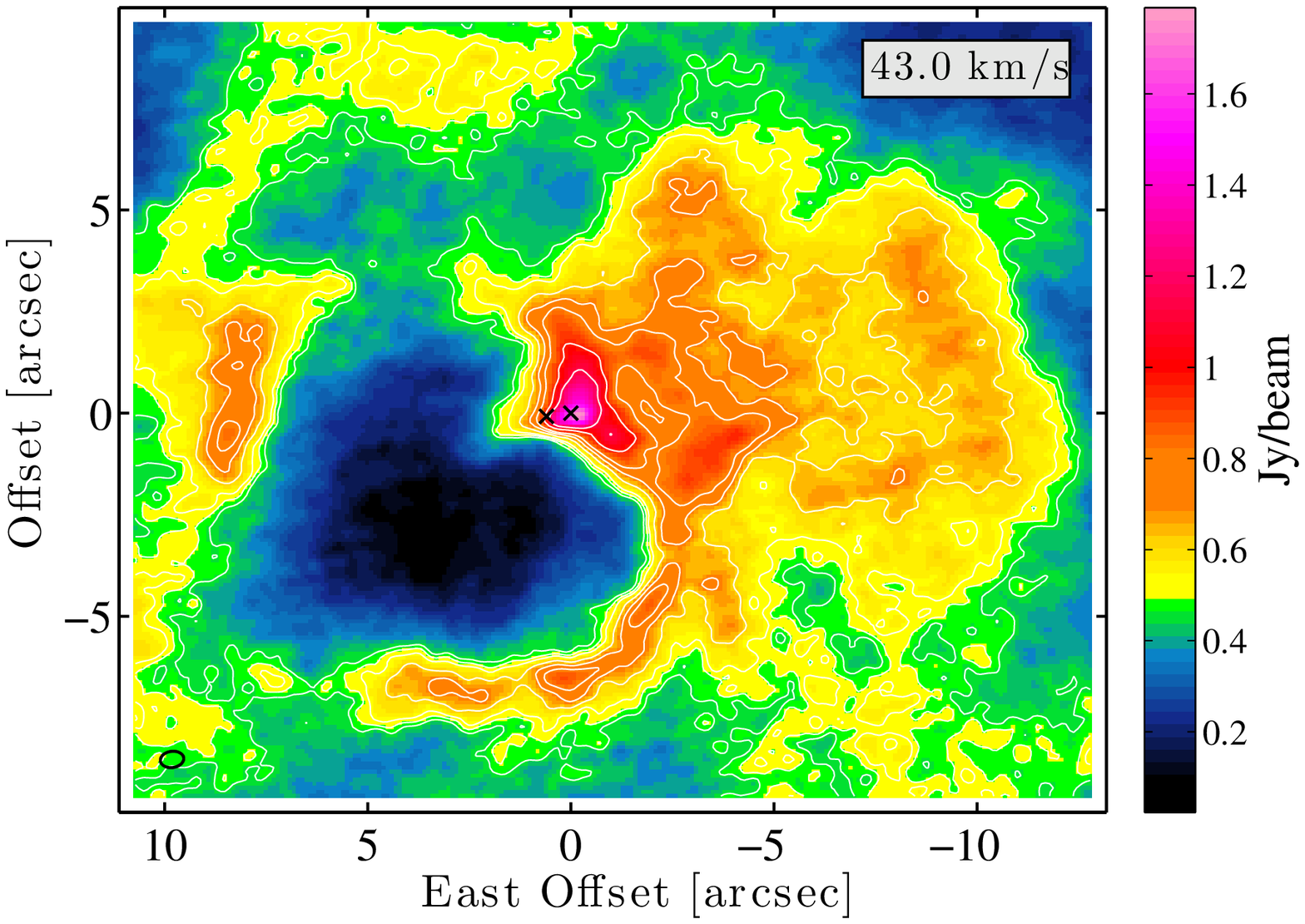} 
   \includegraphics[height=7.5cm]{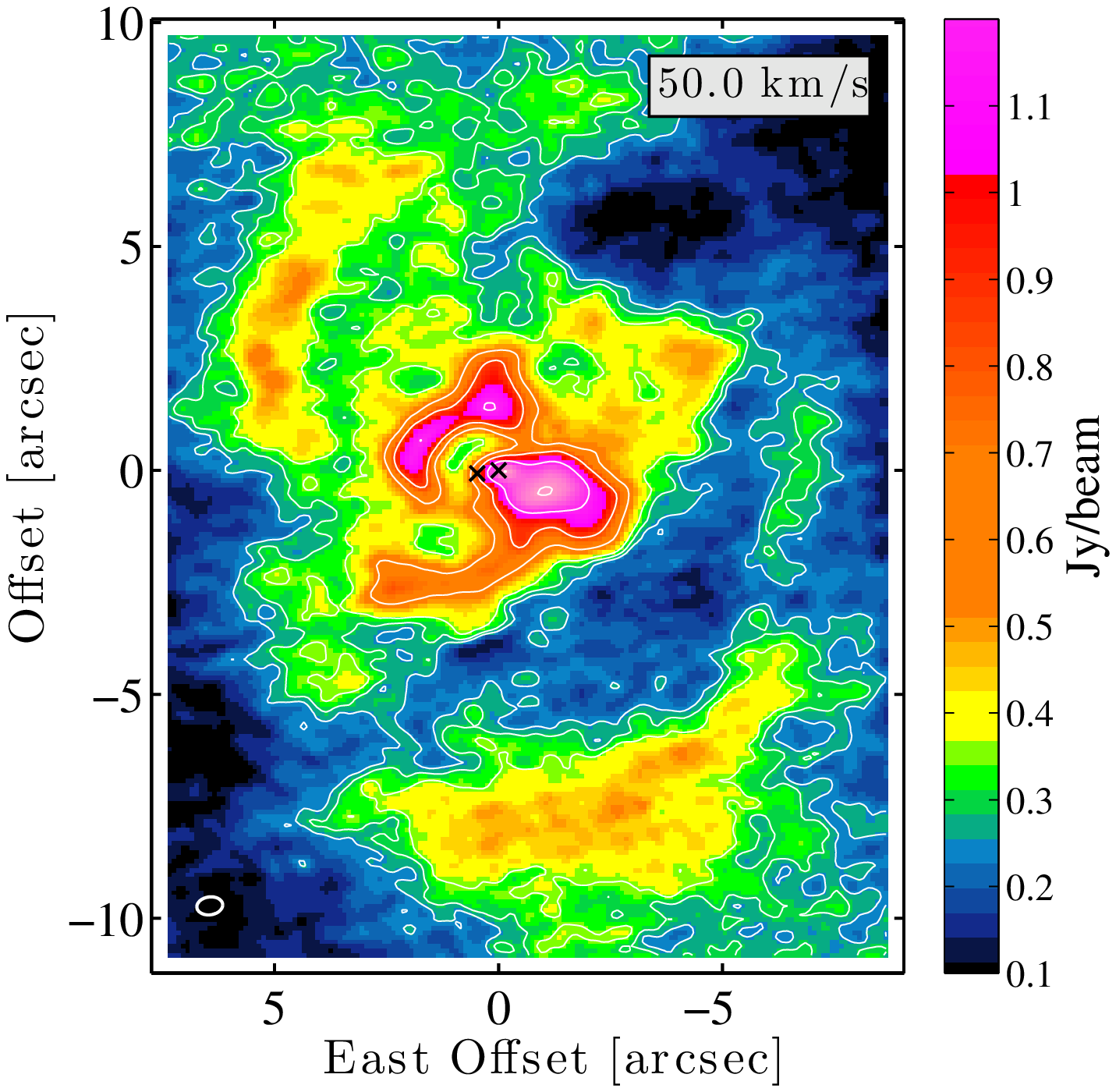} 
   \caption{The CO(3-2) map averaged over 2 and 3\,km\,s$^{-1}$ around $v_{\rm{lsr}}$=43.0 and 50.0\,km\,s$^{-1}$, respectively, showing the bubble structure ({\it left}) and the spiral arcs ({\it right}) discussed in Sect.~\ref{morph}. The crosses mark the position of Mira A (west) and B (east). Contours are drawn at multiples of the rms noise level. The beam is shown in the lower left corner.}
   \label{bubble}
\end{figure*}

The line profile is shown in Fig.~\ref{pv} ({\it left}) and channel maps are presented in Appendix A. Comparison of our spectrum with previous single dish observations \citep{youn95} indicates that even with the ACA only 50\% of the extended flux is recovered. As a result, the channels with strong emission only reach a dynamic range (peak flux/SNR) of $\sim50$. The highly complex circumstellar morphology seen in the channel maps (Fig.~\ref{map}) has clearly been shaped by many different, likely interacting, dynamical processes during the evolution of the system. Here we focus on the two most striking features: the large bubble-like structure found in the blue bump of the line (at $\sim$41.5\,km\,s$^{-1}$, Figs~\ref{pv} and \ref{bubble}, {\it left}), and the spiral arcs  (Fig.~\ref{bubble}, {\it right}) found in the main line component centered on the stellar velocity at 47.0\,km\,s$^{-1}$. 

In the channel maps corresponding to the blue bump in the line profile, there is a large (5-10\arcsec) arc centered on Mira A and facing the companion. It grows with velocity and bends inward at the edges until, at $v_{\rm{lsr}}$=43.0\,km\,s$^{-1}$, the gas forms a distinct bubble (Fig.~\ref{bubble}, {\it left}) on the south-east side of Mira A. At larger velocities, the south part becomes more noticeable until, beyond $v_{\rm{lsr}}$=44.5\,km\,s$^{-1}$, other structures become more apparent. The edges of the bubble are sharp on the east (companion) side of Mira A and no emission above the noise level of the observations is picked up inside the bubble; however, smooth, extended emission is resolved out (also when the ACA is used). 
Figure~\ref{pv} ({\it right}) shows a position-velocity diagram created along the major axis (position angle 38$^{\circ}$) of the south-east bubble. It shows the slight velocity shift between the north and south part and that, e.g., the north part of the arc shifts from 2\arcsec to 6\arcsec offset from 39 to 44\,km\,s$^{-1}$, indicating that part of the frontside of the bubble is visible at the velocities moving toward us. Although low surface brightness prevents detection at the extreme velocities at the very front- and back-side of the bubble, this gives a lower limit of the radial expansion velocity of $\sim$5\,km\,s$^{-1}$. 

Spiral arcs are seen from $v_{\rm{lsr}}$=45 to 54\,km\,s$^{-1}$ (Fig.~\ref{map}). These channels make up the main part of the CO line profile and dominate the emission. At $v_{\rm{lsr}}$=47.5\,km\,s$^{-1}$, the accretion wake behind the companion is apparent and it is present through all redder velocity channels (Figs~\ref{bubble} and \ref{spiralwake}). The spiral can be traced out to 2-3 windings (Fig.~\ref{spiralwake}) and different parts are visible in different velocity channels indicating that it is flatter and more confined to the orbital plane than the spiral around, e.g., R~Scl \citep{maeretal12}. Since the orbital plane is inclined and seen close to edge on (22$^{\circ}$ south of edge on), the real winding separation cannot be determined directly from the map. The projected separation is of the order of 3\arcsec--5\arcsec, which would correspond to 8\arcsec--13\arcsec, if the spiral is confined to the orbital plane.

\section{Discussion}
\label{dis}

\subsection{Comparison to previous observations}
\label{comp}
From the continuum emission (Sect.~\ref{sep}), the current position of Mira B is $\approx$0\farcs5~east of Mira A at a position angle of 98$^{\circ}$. This agrees with predictions from the orbit by \citet{prieetal02}. 

The same north-south asymmetry, as seen in previous CO observations, is found in the ALMA CO line data; however, due to the greatly improved spatial resolution and sensitivity, the ALMA data shows much more detail. For instance, it is likely that the north-east feature at 42.5\,km\,s$^{-1}$ in the OVRO map \citep{planetal90}, is the large blue-shifted arc detected with ALMA, but at low resolution. The butterfly shape found in the PdB map \citep{jossetal00} is not apparent in the ALMA data, but it is possible that it is a low-resolution version of the structures seen in Fig.~\ref{bubble} ({\it left}). In the ALMA maps (Fig.~\ref{map}) there is no clear indication of the molecular gas being disrupted by a high velocity outflow, as the one detected in KI emission \citep{jossetal00}, or by the north and south stream. 

It is not straightforward to connect the large dust arcs found in the PACS image \citep{mayeetal11}, and the inner spiral shape found in the ALMA data, since the dust distribution is affected by other processes. Also, the PACS image only shows the velocity-integrated information and detailed structures may not be apparent \citep[cf. PolCor and ALMA images of the spiral and shell around R~Scl,][]{maeretal14}. The separation between the dust arcs is of the order 15\arcsec-30\arcsec~which is significantly larger than in the inner spiral detected with ALMA. 

Finally, the interaction between the circumstellar material and the ISM is seen in the GALEX images \citep{martetal07}, and a large bow shock is detected $\approx$5\arcmin~south of Mira AB. While the large-scale dust distribution is clearly shaped by this interaction \citep{mayeetal11}, this is not the case for the smaller-scale gas distribution mapped by ALMA. The presence of the bow shock much further out means that the inner structures cannot be explained by (hydrodynamic) interaction with the ISM. 

\subsection{Possible shaping scenarios}
\label{shape}
The bubble found on the south-east side of Mira A is $\sim$800\,AU across, and there is a similar, filled structure on the opposing, north-west side of the star (Sect.~\ref{morph} and Fig.~\ref{bubble}, {\it left}). One possible explanation for these  shapes is that the wind from Mira B is blowing a hole in the circumstellar material from Mira A. The time scale for a continuous wind \citep[assuming the larger Mira B mass-loss rate, 10$^{-11}$\,M$_{\odot}$\,yr$^{-1}$, and wind velocity, 450\,km\,s$^{-1}$,][]{woodkaro06} to create a bubble of that size is $\sim$380 years (assuming $\dot{M}\sim$10$^{-7}$\,M$_{\odot}$\,yr$^{-1}$ from Mira A). The terminal velocity of the shell surrounding the bubble would be $\approx$5\,km\,s$^{-1}$ \citep[e.g., the wind-wind interaction model for planetary nebulae,][]{kwok83}. The bubble is not centered on Mira B because, in the direction of Mira A, the CSE is too dense to be significantly affected by Mira B's wind. The wind will predominantly blow in the north-south direction since it is hindered by the (almost edge-on) dense accretion disk around Mira B, and this could explain the bi-polar shape of the bubble. This scenario is consistent with the shell radial velocity of $\gtrsim$5km\,s$^{-1}$ (Fig.~\ref{pv}, {\it right}), and the mirrored shape on the opposite side of Mira A, that could have been created half a period ago. The north-west bubble seems to be almost filled with material from the wind of Mira A, which agrees with the time scales and an expansion velocity of 5\,km\,s$^{-1}$; however the time scale for the creation of the bubble from these order-of-magnitude estimates (assuming spherically symmetric winds) is slightly longer than the time Mira B would stay on either side of Mira A if the orbital period is 500 yrs.

The two companions seem surrounded by a common dense molecular envelope, and at red-shifted velocities, there appears to be an accretion wake behind Mira B. Several arcs centered on Mira A are found in the line emission from 45.5 to 54.5\,km\,s$^{-1}$ (Figs~\ref{bubble}, \ref{map} and \ref{spiralwake}). Even though these arcs are physically connected in a continuous spiral, they will appear as separate arcs in velocity space, since different sections of the spiral arms will have different velocities along the line of sight. The relatively low wind velocity of Mira A ($\sim$5\,km\,s$^{-1}$) will cause the circumstellar material to slowly fill its Roche lobe and eventually fall onto Mira B through the inner Lagrangian point \citep[Wind Roche-lobe Overflow (WRLOF),][]{mohapods07}. This will focus the outflow towards the orbital plane and the orbital motion will result in a rather flat spiral \citep{mohapods12}, unlike the case of, e.g.,~R~Scl \citep{maeretal12}, where the AGB wind has a much higher expansion velocity (14.5\,km\,s$^{-1}$) resulting in a spiral with greater vertical extension. Detailed modelling is beyond the scope of this paper and will be presented in a future publication. Preliminary results show that the spiral spacing will be on the order of 500--1000\,AU or roughly 5\arcsec--10\arcsec~for a distance of 100 pc (assuming the gas outflow velocity of 5\,km\,s$^{-1}$ and orbital period between 500 yrs and 1000 yrs). The upper end of these theoretical estimates fit well within the observed spiral separation when we allow for orbital inclination effects.

\section{Conclusions}
The CO(3-2) emission from the Mira AB system has been mapped with ALMA in Cycle 1. The maps show the circumstellar molecular gas distribution in amazing detail with the highest spatial resolution achieved so far. The complex morphology confirms that circumstellar material has been shaped by several interacting dynamical processes (e.g., mass loss, binary shaping, wind-wind interaction, accretion) during the evolution of the star, and the data presented here can be analysed to better understand these processes.

The binary pair is marginally resolved in the continuum and has a separation of $\approx$0\farcs5~at a position angle of 98$^{\circ}$. We suggest that the features seen in the CO gas emission confirms the scenario where the slow wind of the AGB star fills its Roche lobe and flows onto the companion in the orbital plane \citep{mohapods07}. A relatively flat spiral is formed and seen as separate arcs at different velocities. The spacing between the spiral arms agrees with what is expected from the orbit and wind properties of Mira A. The accretion of material onto the companion gives rise to a varying wind already detected in UV line emission \citep{woodkaro04}. A tentative explanation is that this wind blows a large bubble in the circumstellar material as seen at slightly blueshifted velocities in the ALMA data. The wind from Mira B is pinched by the accretion disk is therefore oriented preferentially in the north-south direction. Its flow is also confined by the dense wind of Mira A to the east. Our estimate of the radial velocity of the material in the shell around the bubble fits this scenario, while the time scale for the creation of the bubble is somewhat longer than the orbital period.

\begin{acknowledgements}
The authors would like to thank the staff of the Nordic ALMA ARC node for their indispensable help and support. This paper makes use of the following ALMA data: ADS/JAO.ALMA\#2012.1.00524.S . ALMA is a partnership of ESO (representing its member states), NSF (USA) and NINS (Japan), together with NRC (Canada) and NSC and ASIAA (Taiwan), in cooperation with the Republic of Chile. The Joint ALMA Observatory is operated by ESO, AUI/NRAO and NAOJ. WV acknowledges support from Marie Curie Career Integration Grant 321691 and ERC consolidator grant 614264.
\end{acknowledgements}


\bibliographystyle{aa}
\bibliography{mira}

\appendix
\section{}
Maps of the CO(3-2) emission from 36.7 to 53.7\,km\,s$^{-1}$ at approximately 0.85\,km\,s$^{-1}$ spectral resolution. As the maps with strong line emission are limited to a dynamic range of $\sim50$ because a significant amount of large scale (>9") flux is not recovered, the rms varies between 30 and 160~mJy in the different velocity channels.
\begin{figure*}[htbp]
   \centering
   \includegraphics[width=4.2cm]{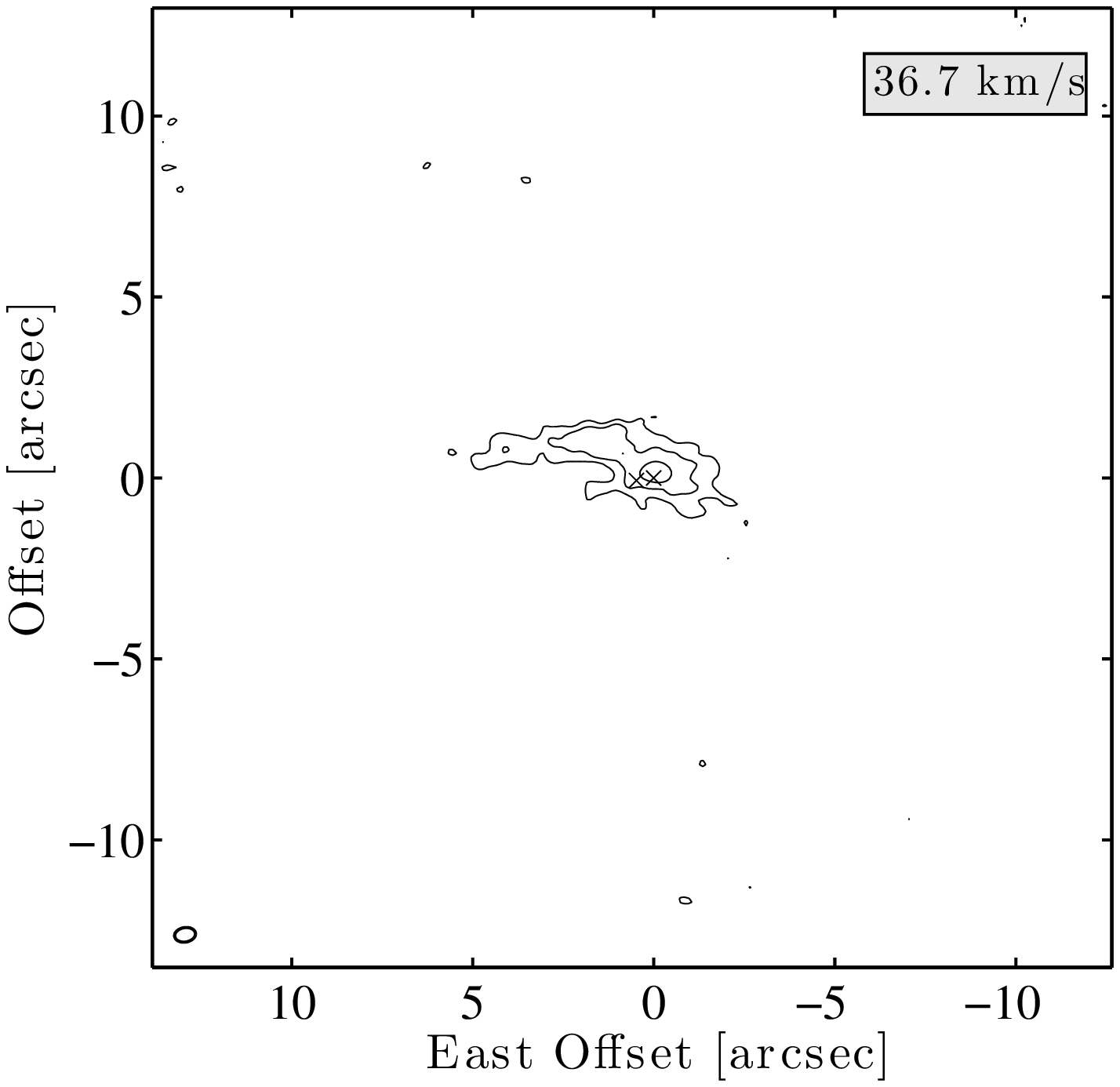} 
   \includegraphics[width=4.2cm]{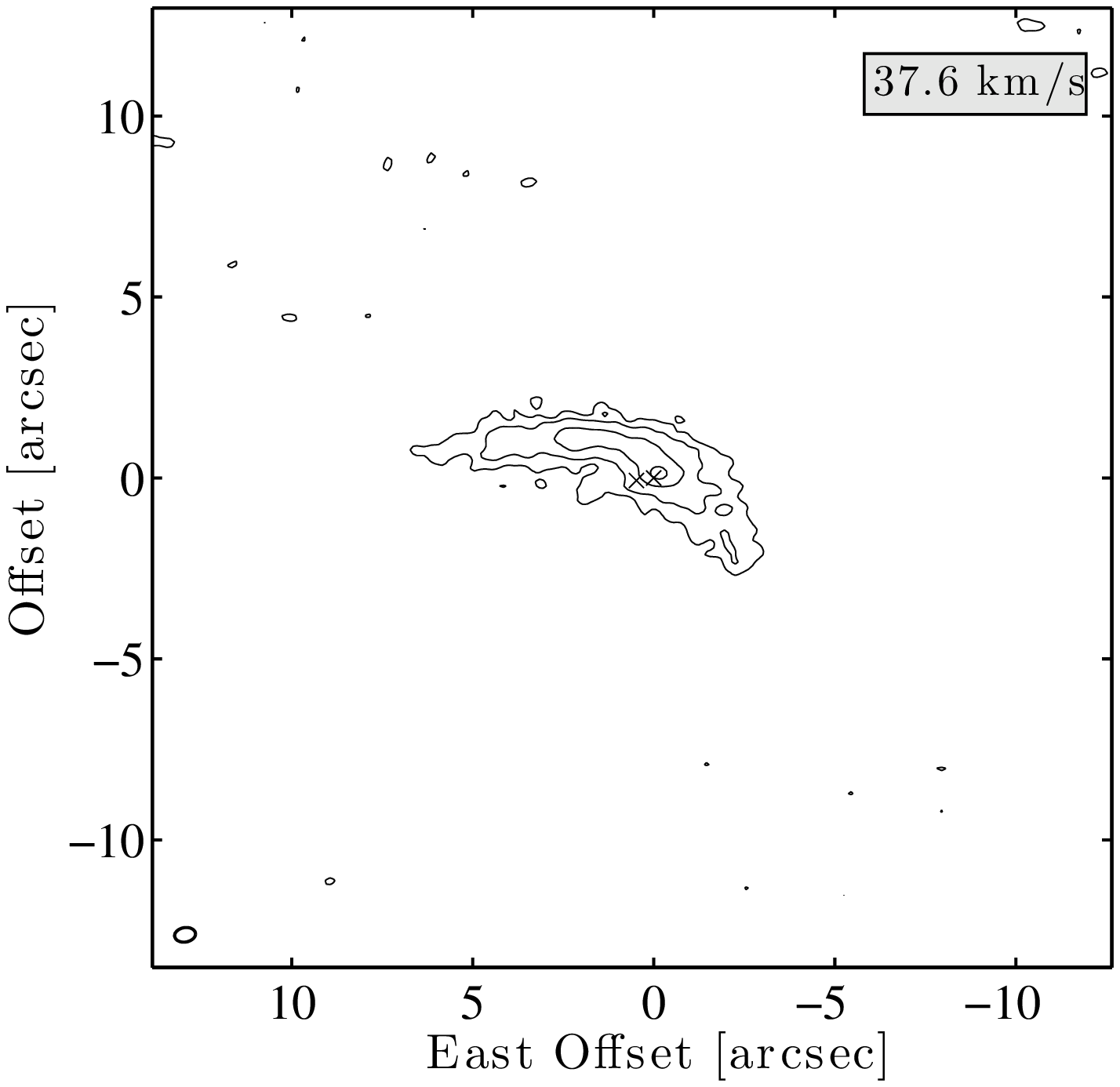} 
   \includegraphics[width=4.2cm]{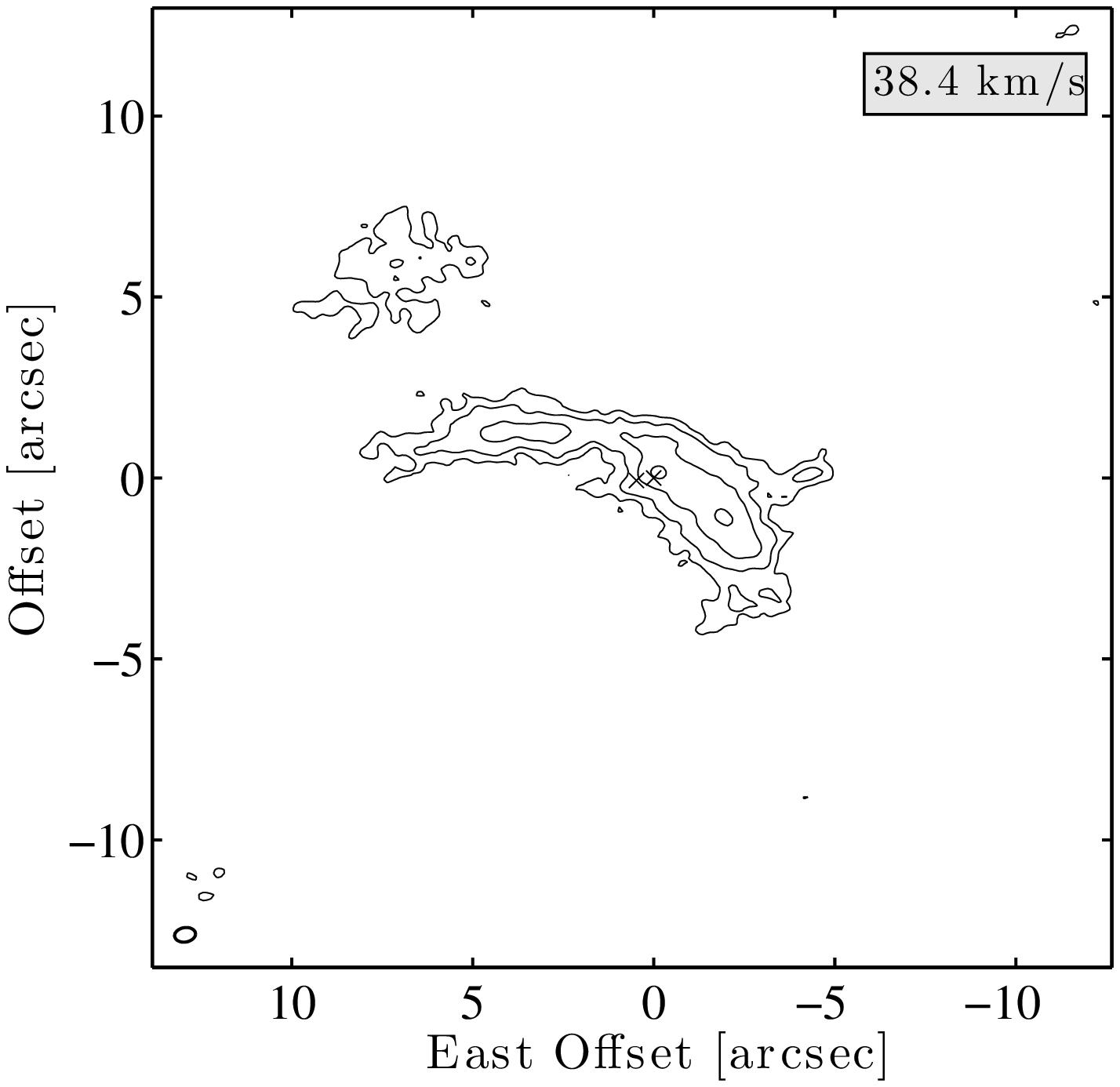} 
   \includegraphics[width=4.2cm]{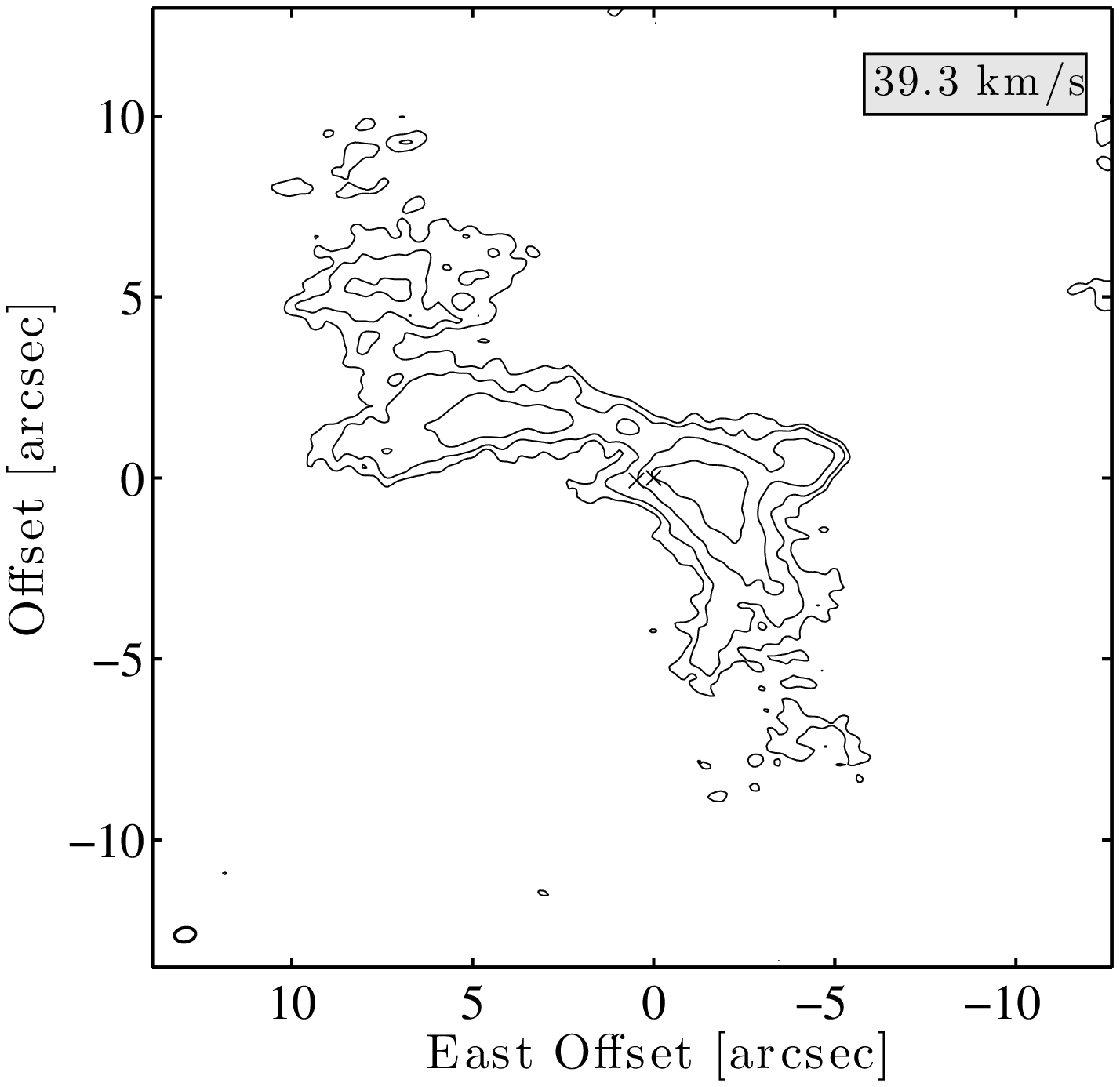} 
   \includegraphics[width=4.2cm]{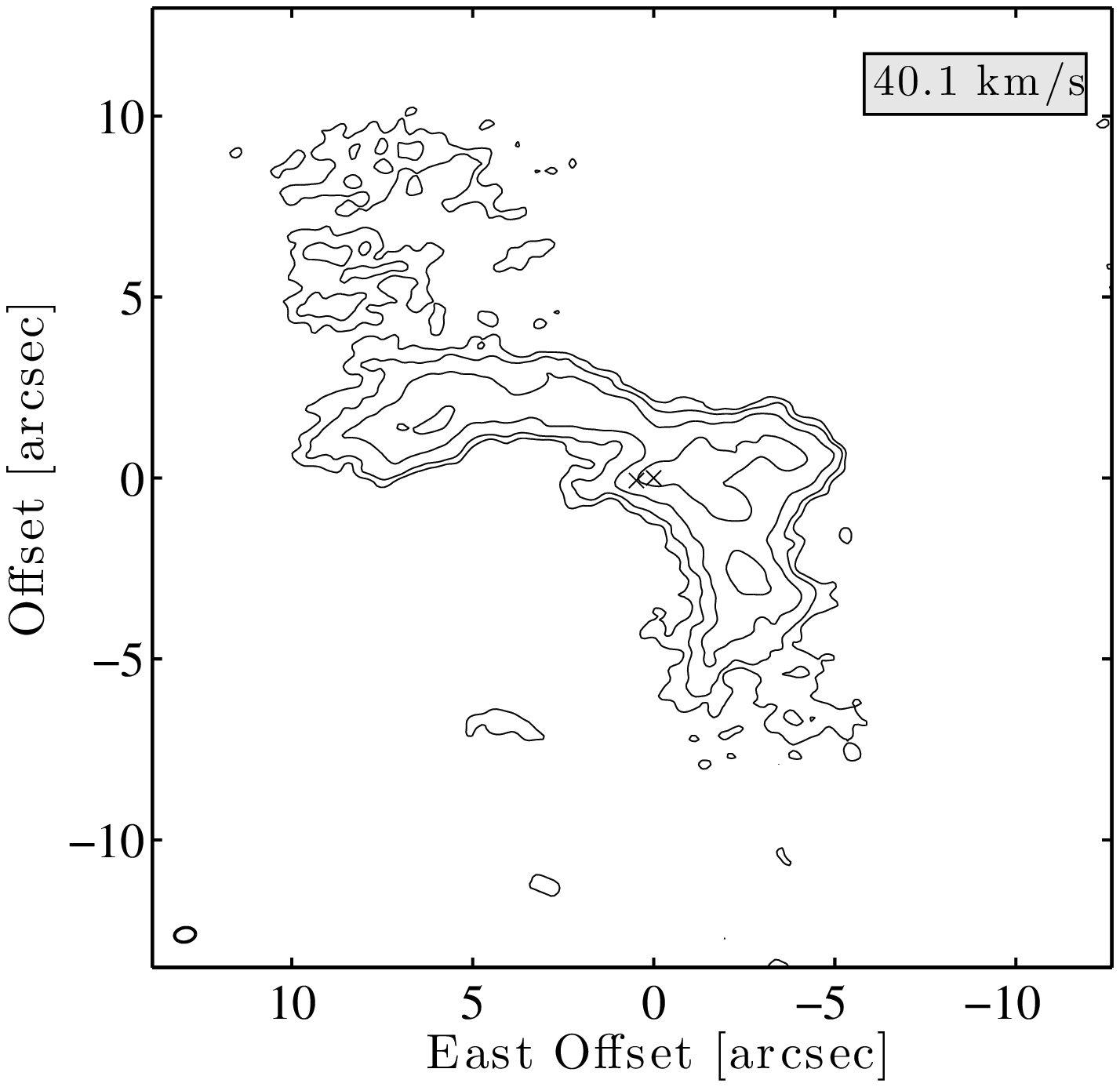} 
   \includegraphics[width=4.2cm]{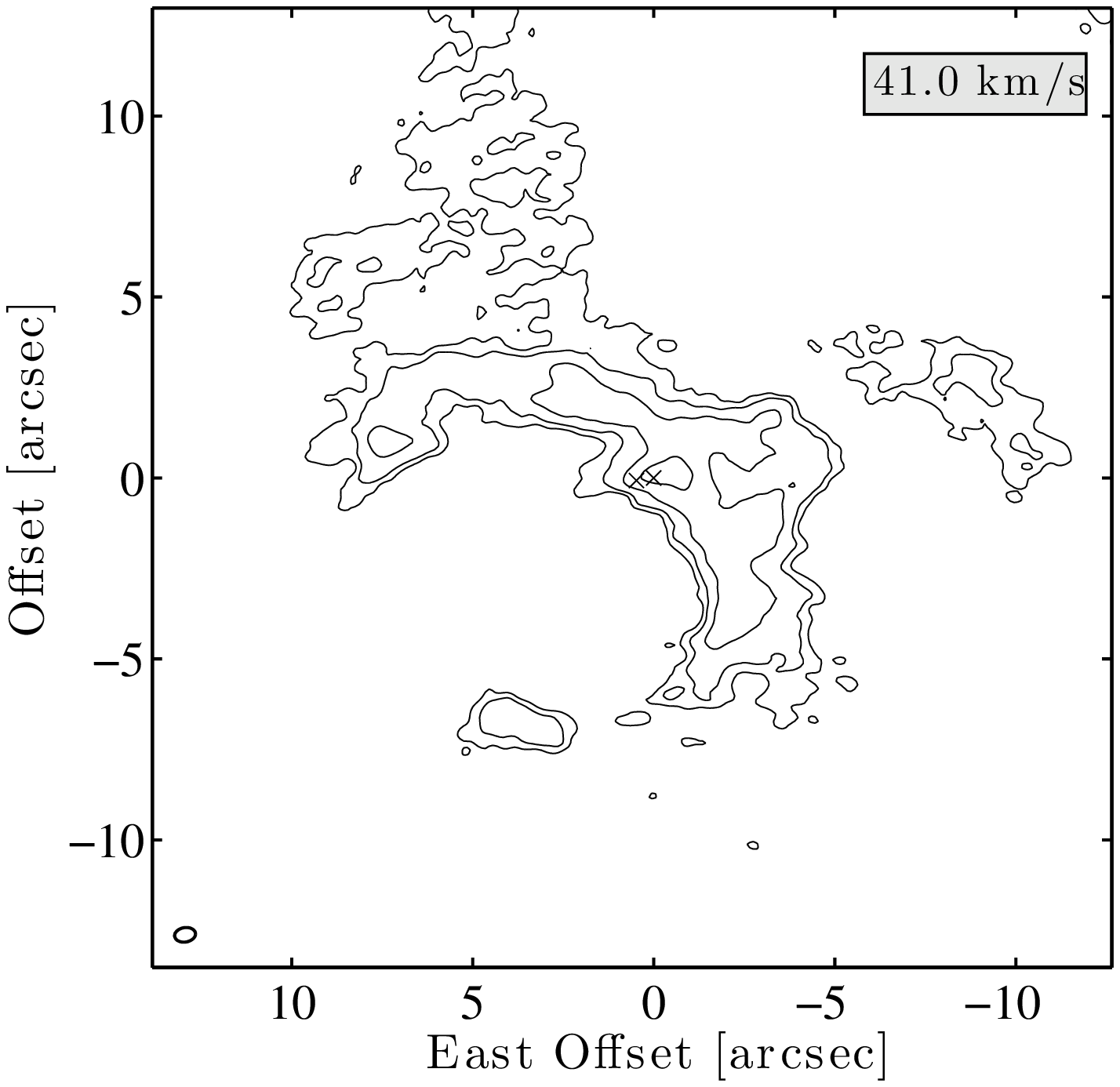} 
   \includegraphics[width=4.2cm]{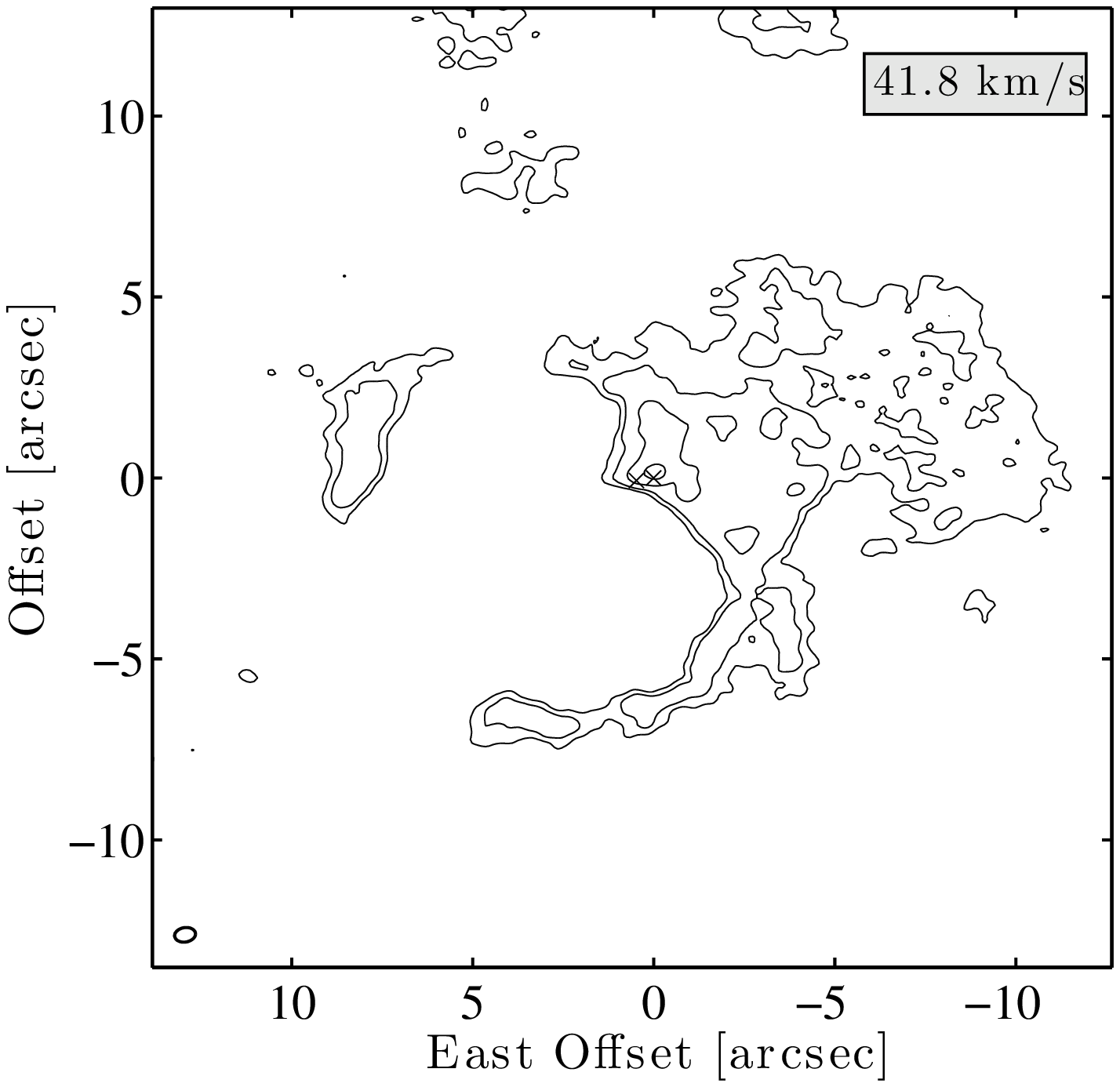} 
   \includegraphics[width=4.2cm]{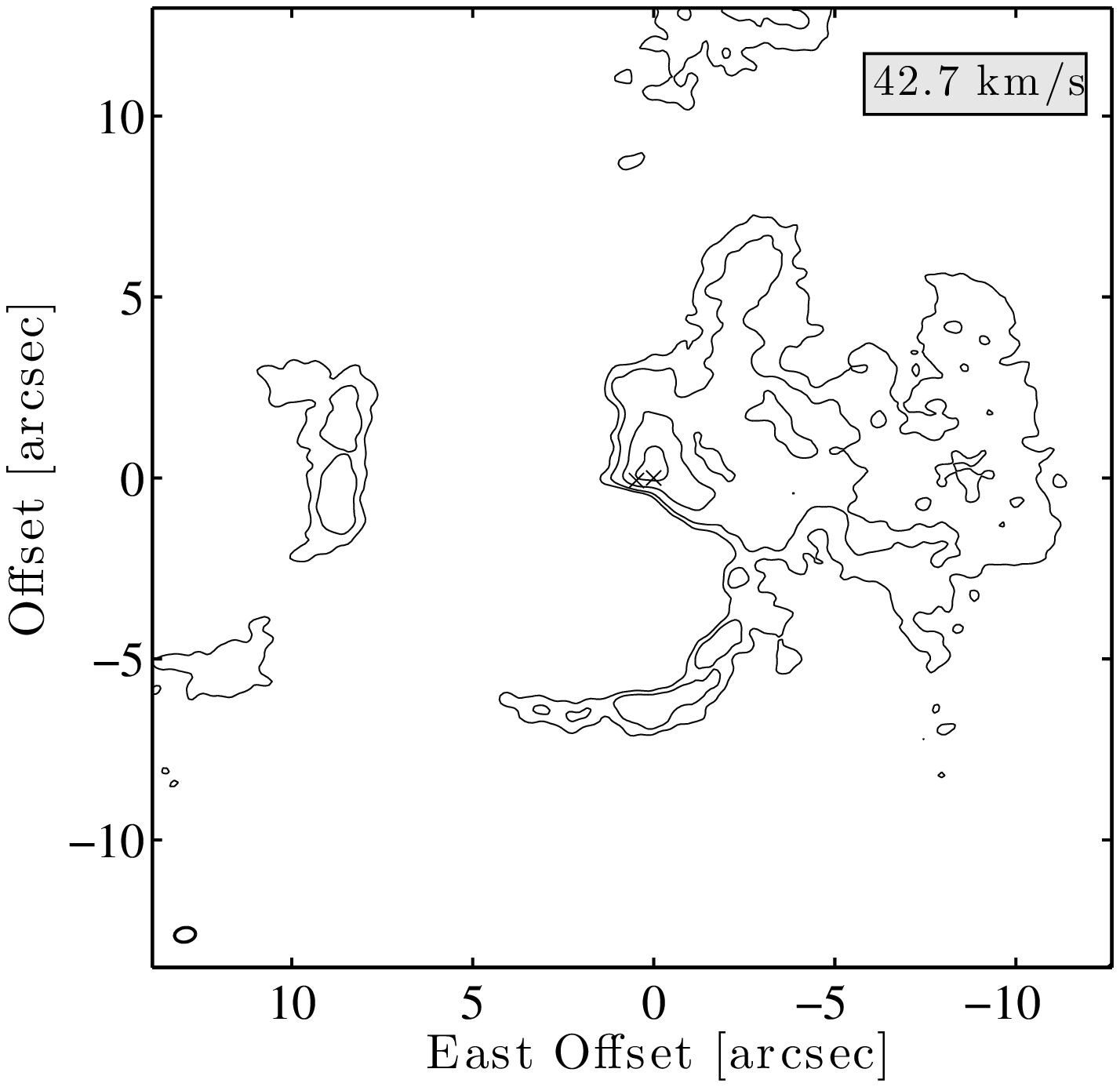} 
   \includegraphics[width=4.2cm]{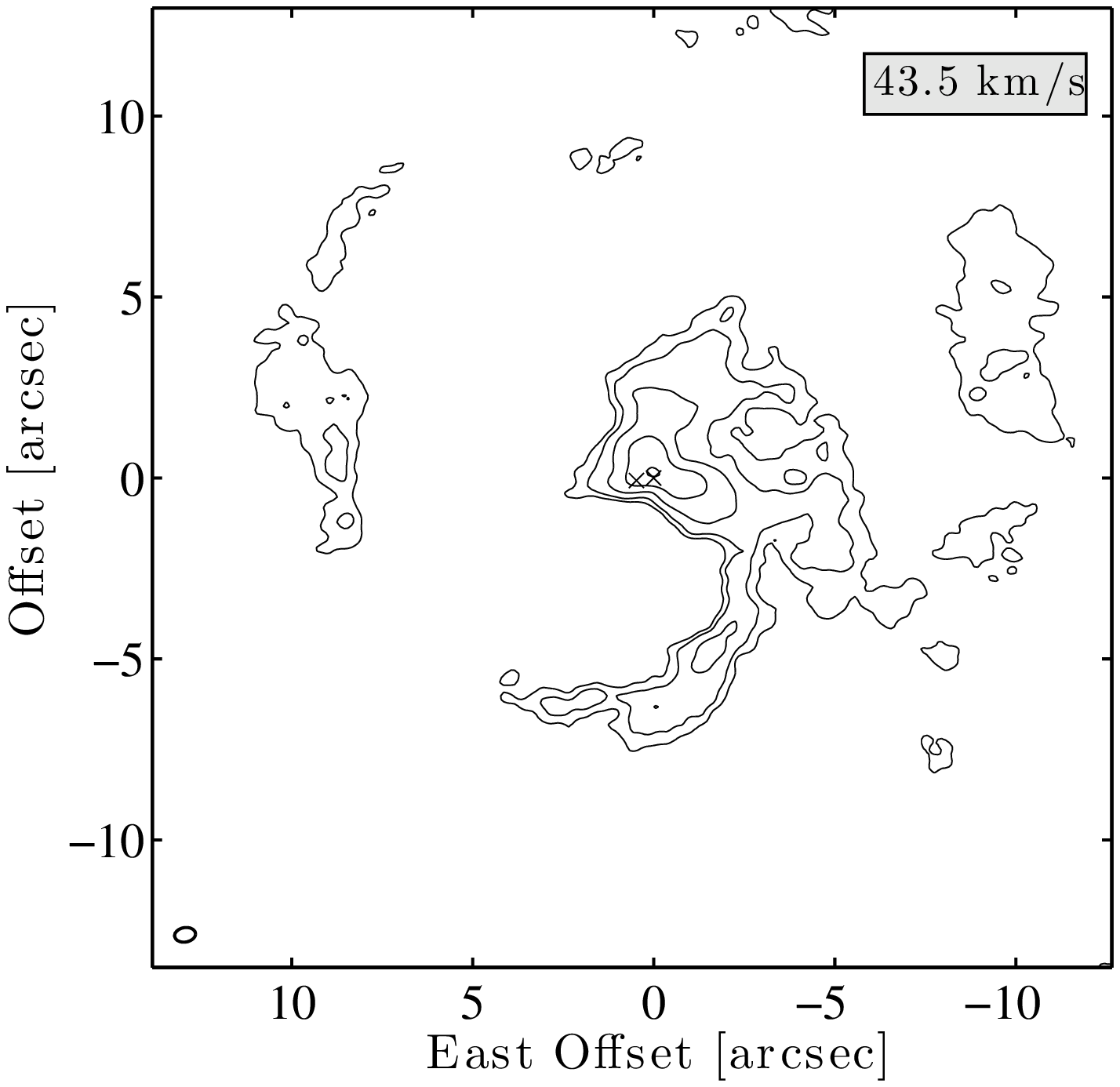} 
   \includegraphics[width=4.2cm]{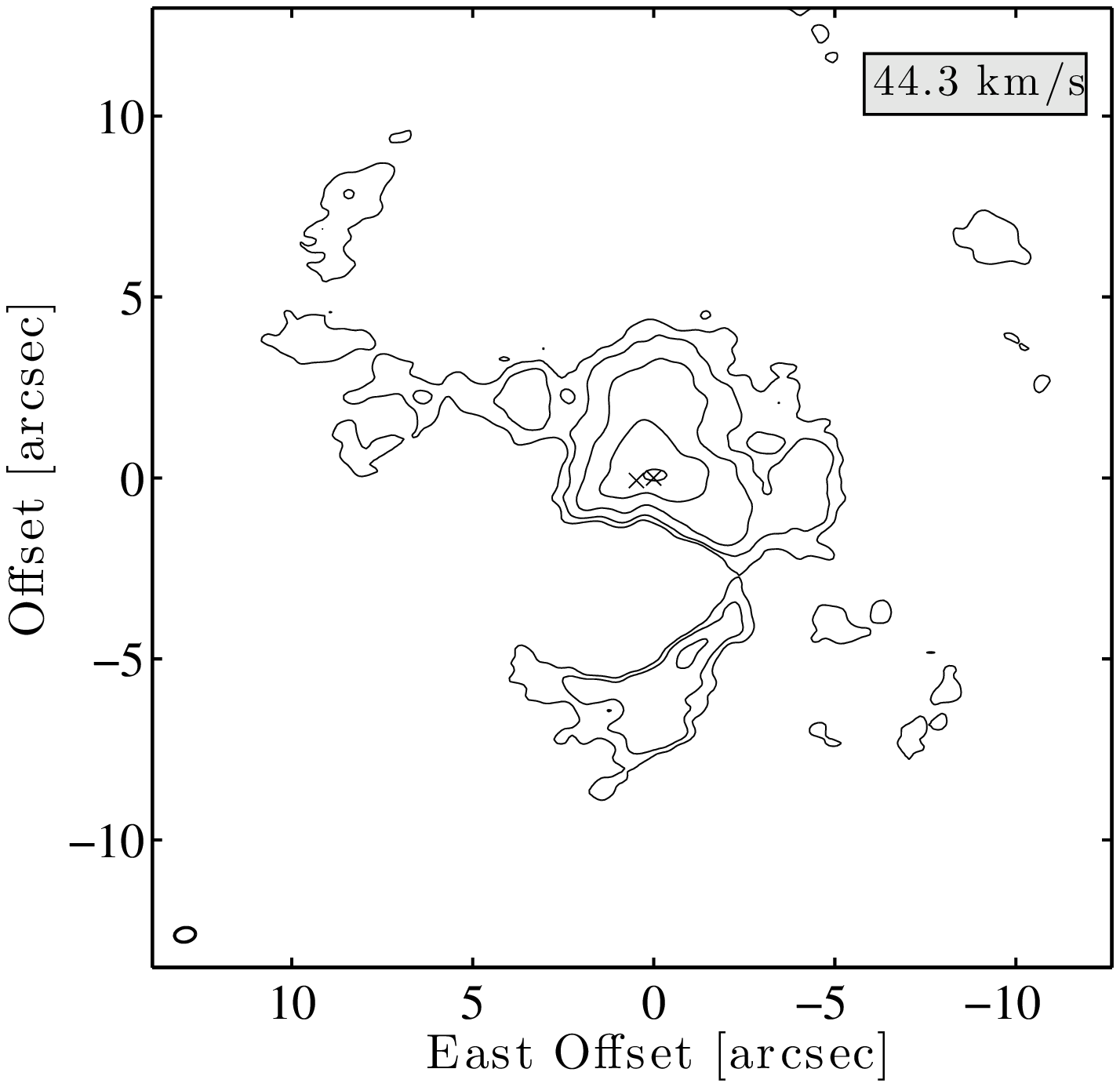} 
   \includegraphics[width=4.2cm]{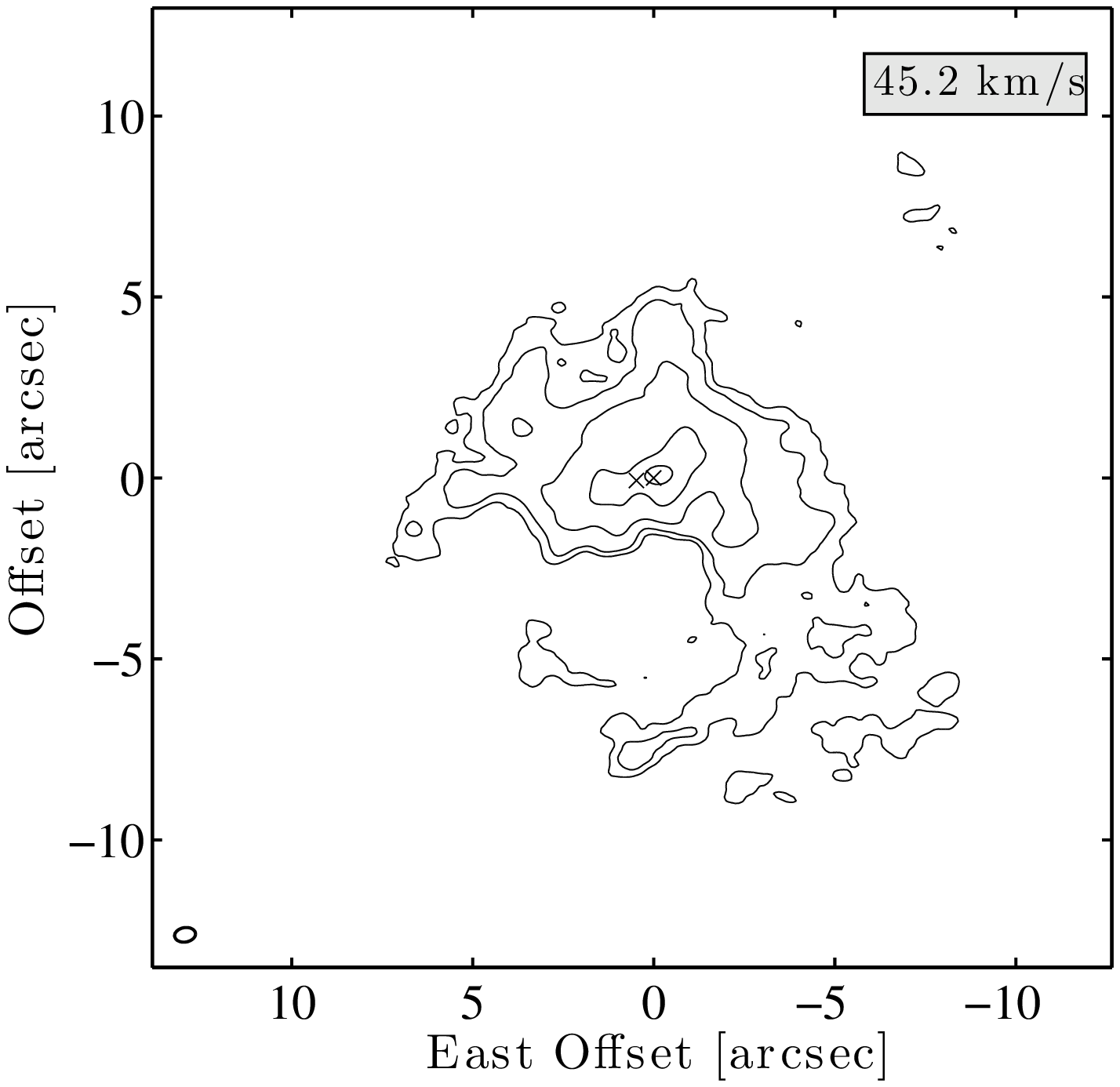} 
   \includegraphics[width=4.2cm]{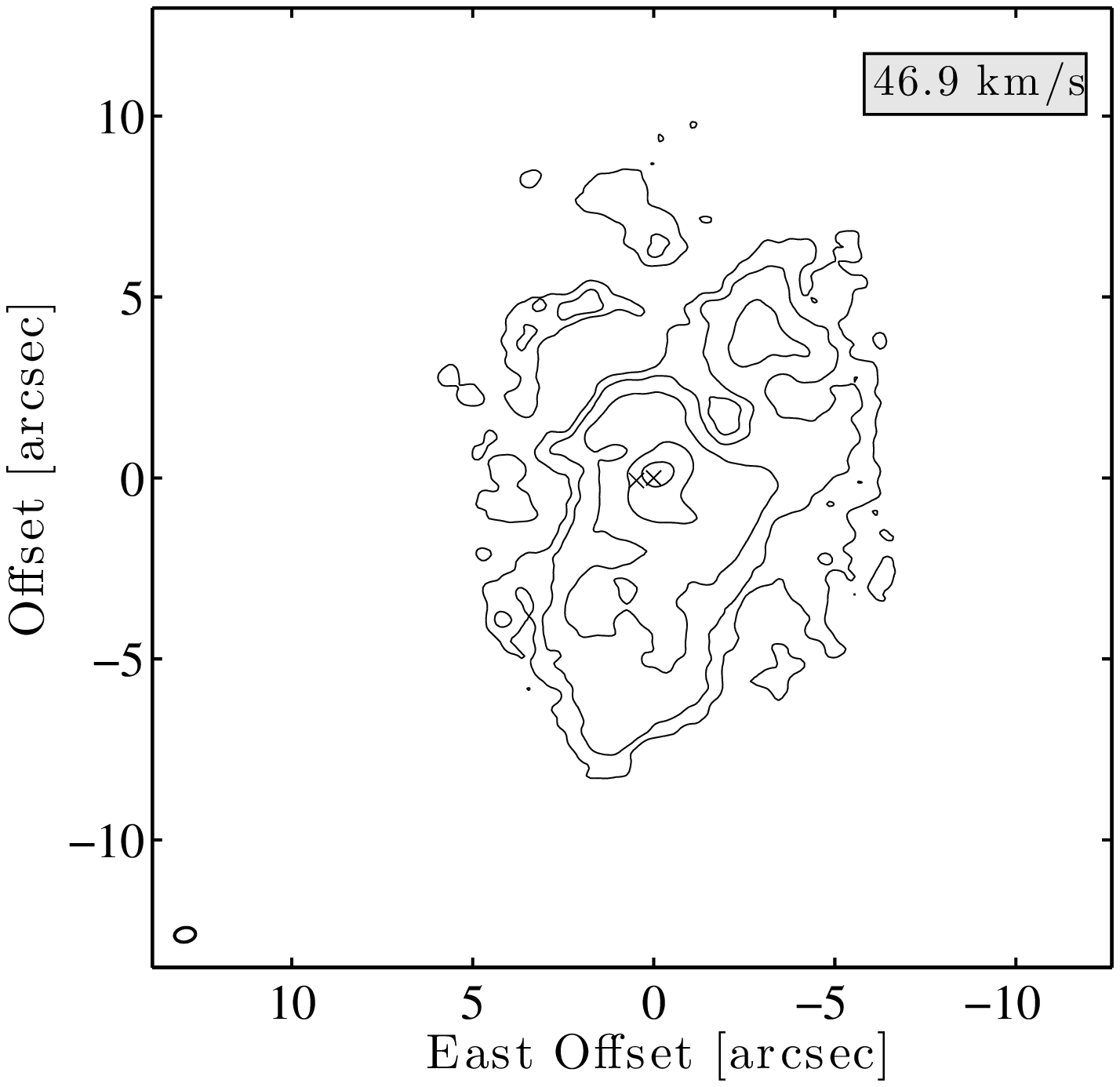} 
   \includegraphics[width=4.2cm]{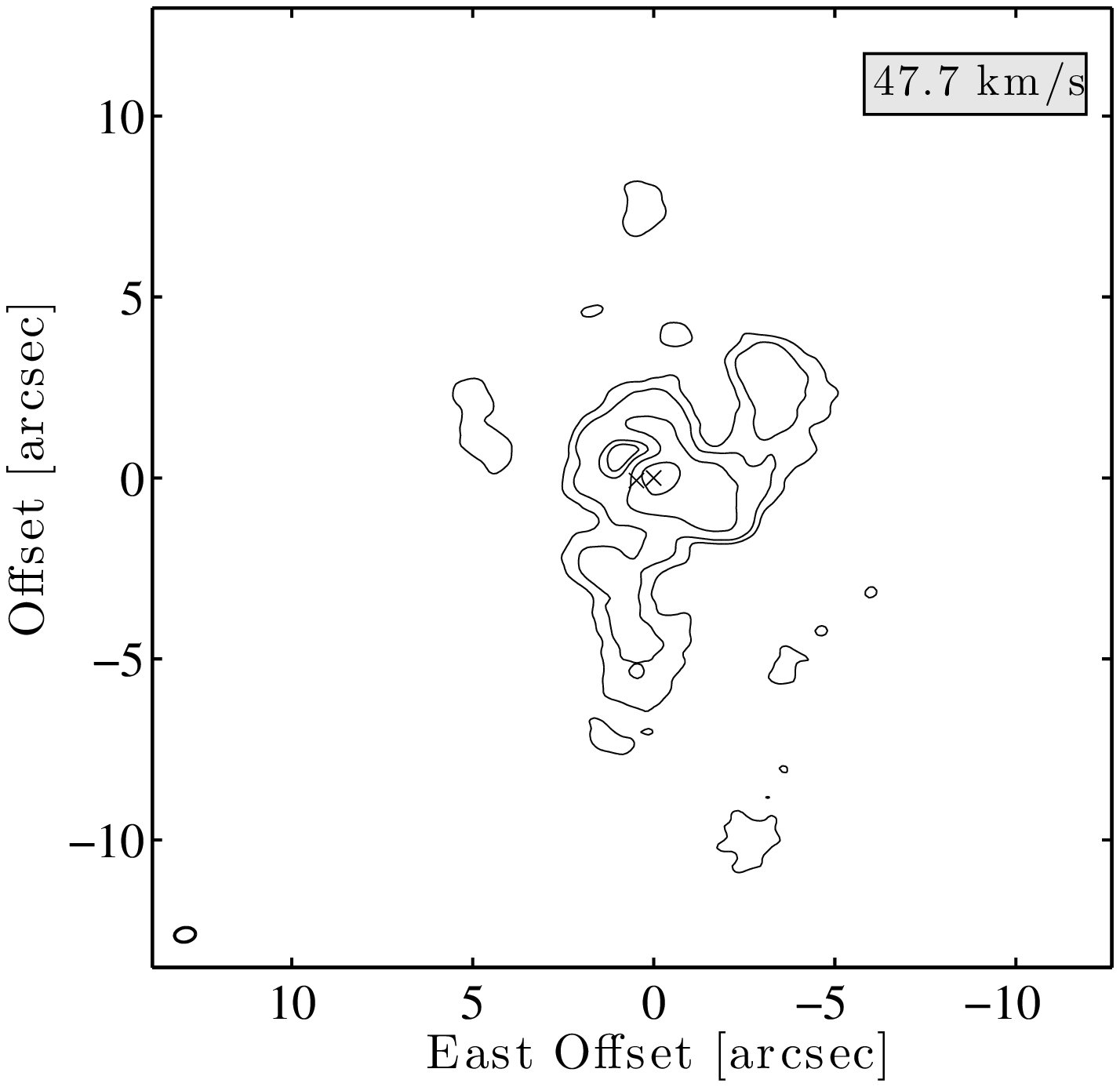} 
   \includegraphics[width=4.2cm]{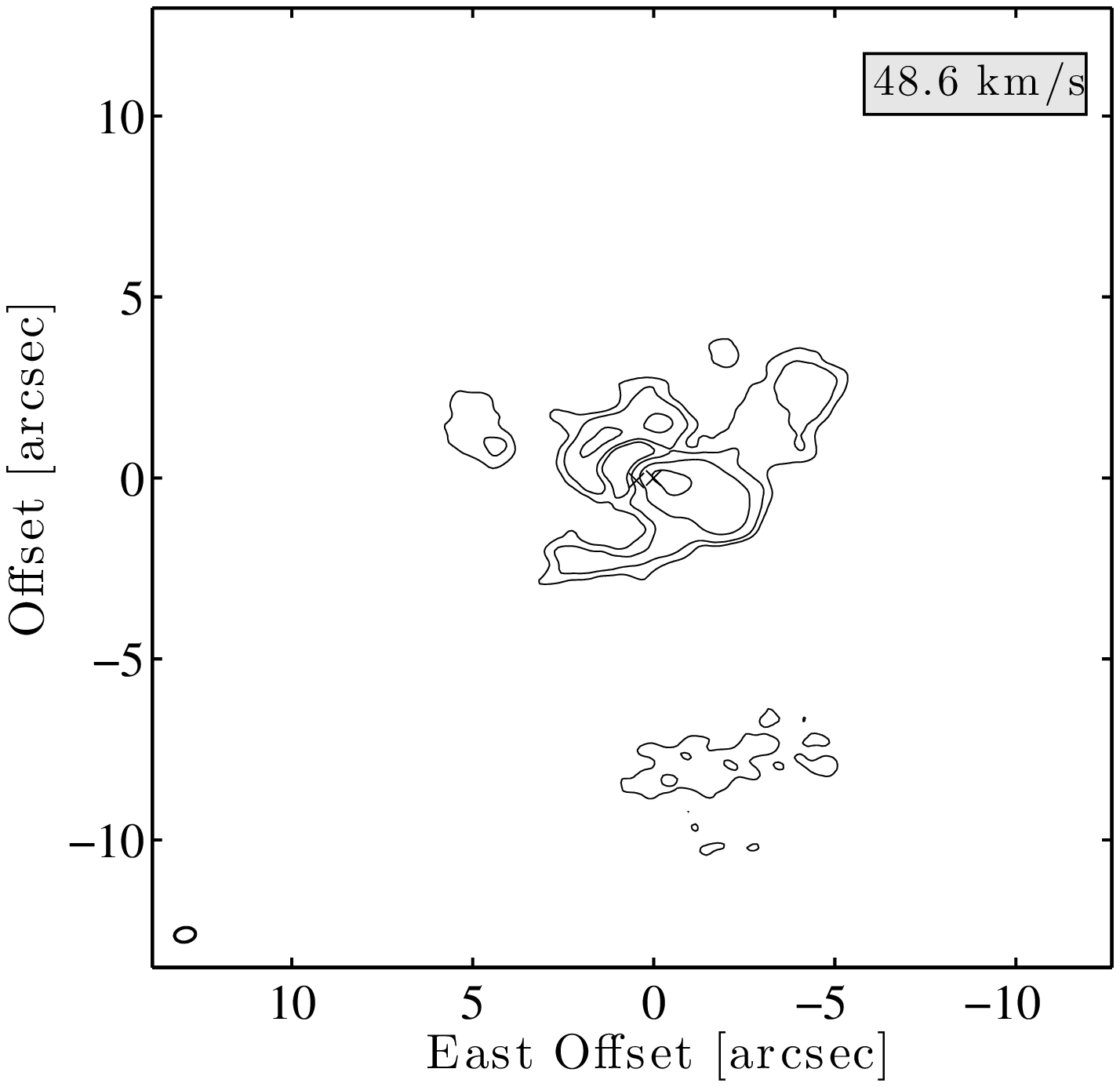} 
   \includegraphics[width=4.2cm]{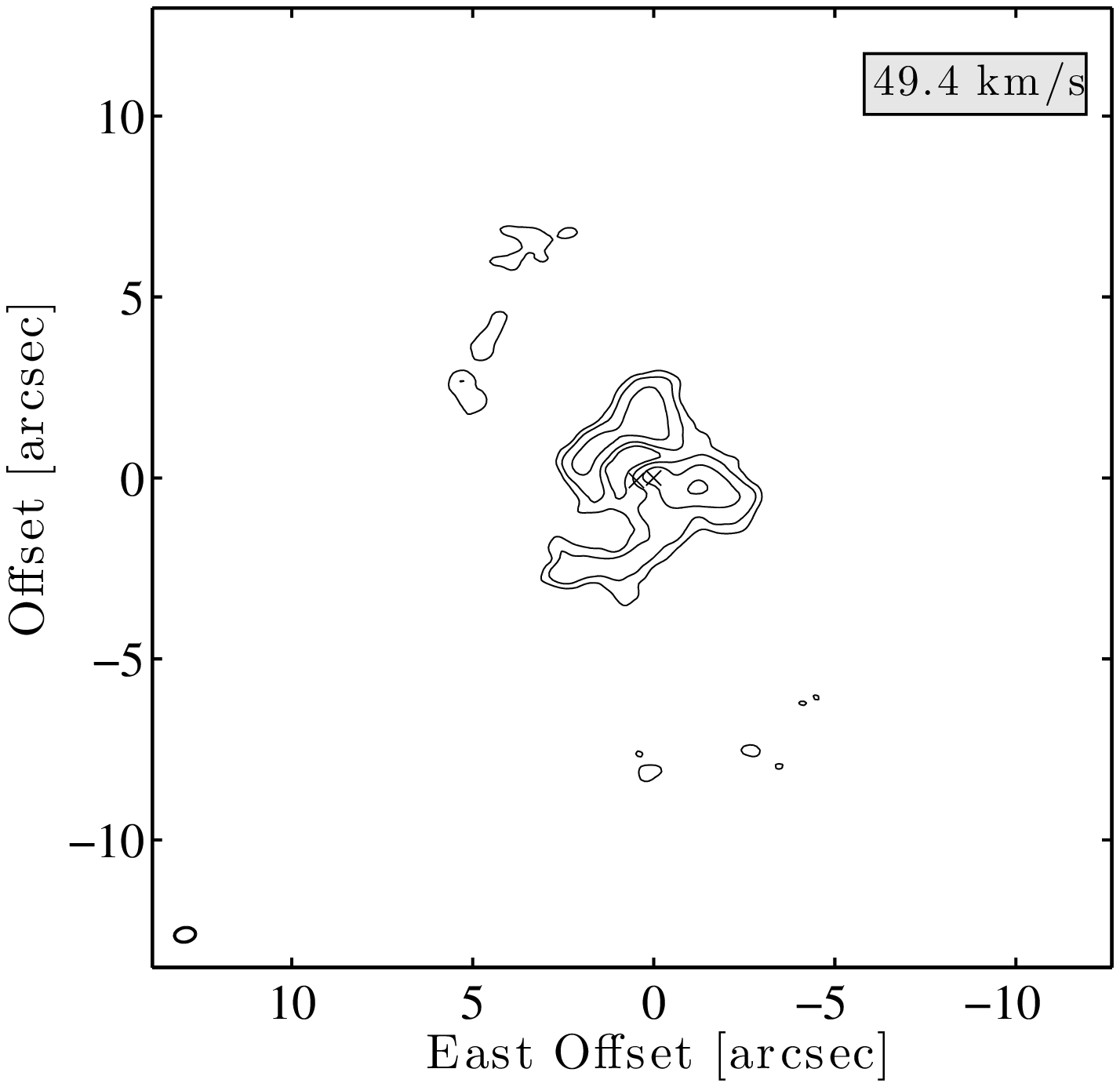} 
   \includegraphics[width=4.2cm]{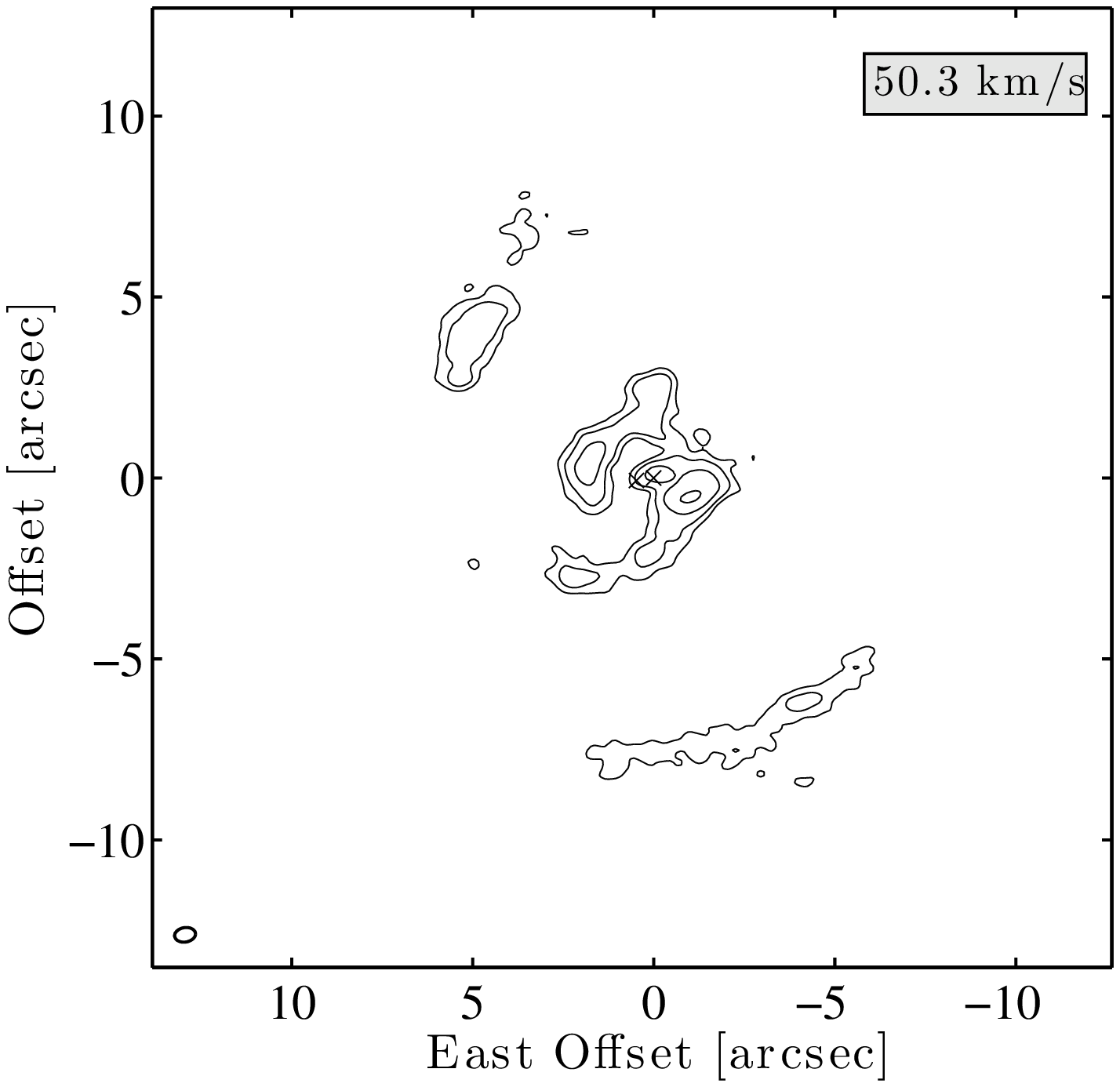}
   \includegraphics[width=4.2cm]{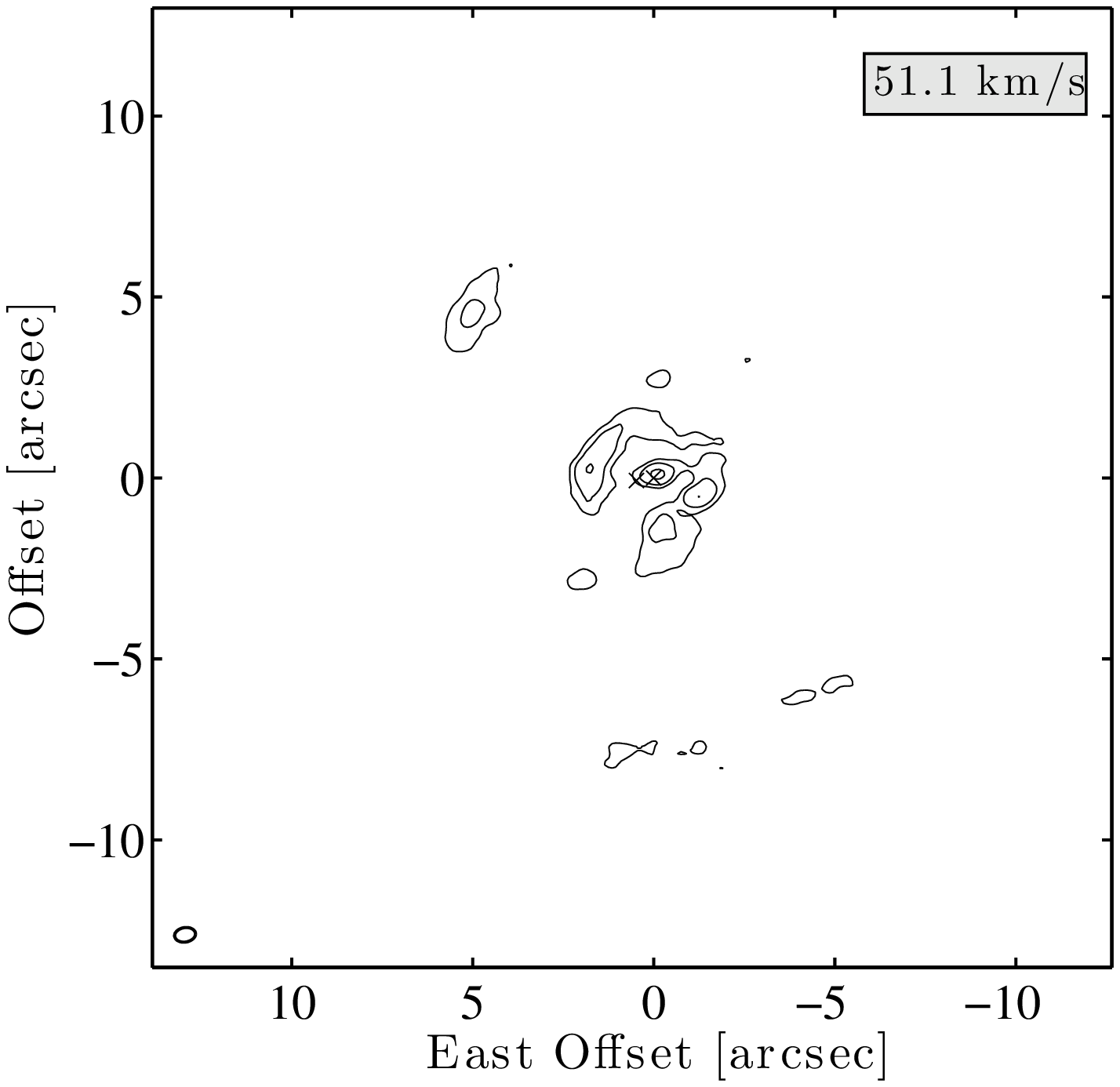} 
   \includegraphics[width=4.2cm]{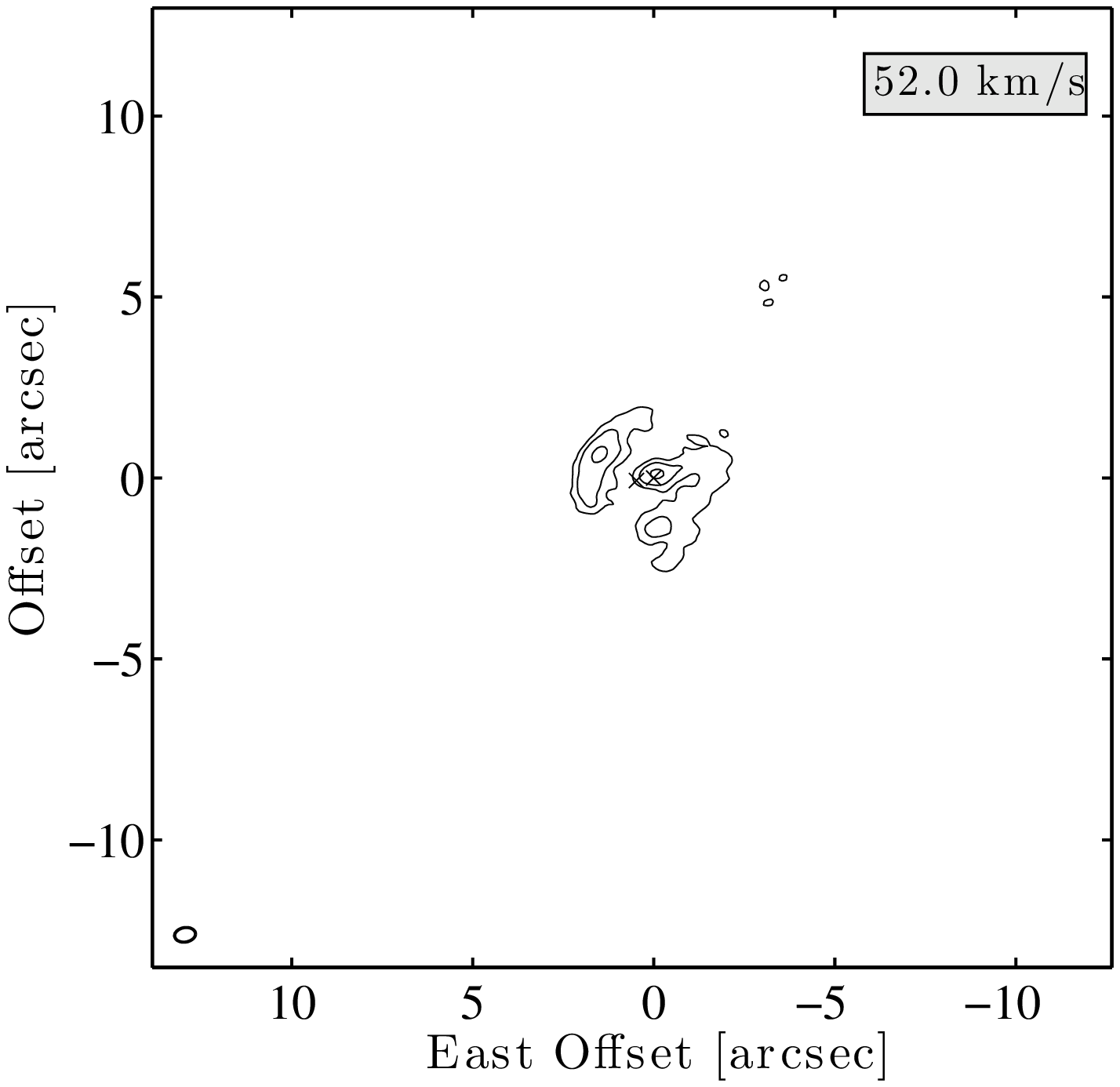} 
   \includegraphics[width=4.2cm]{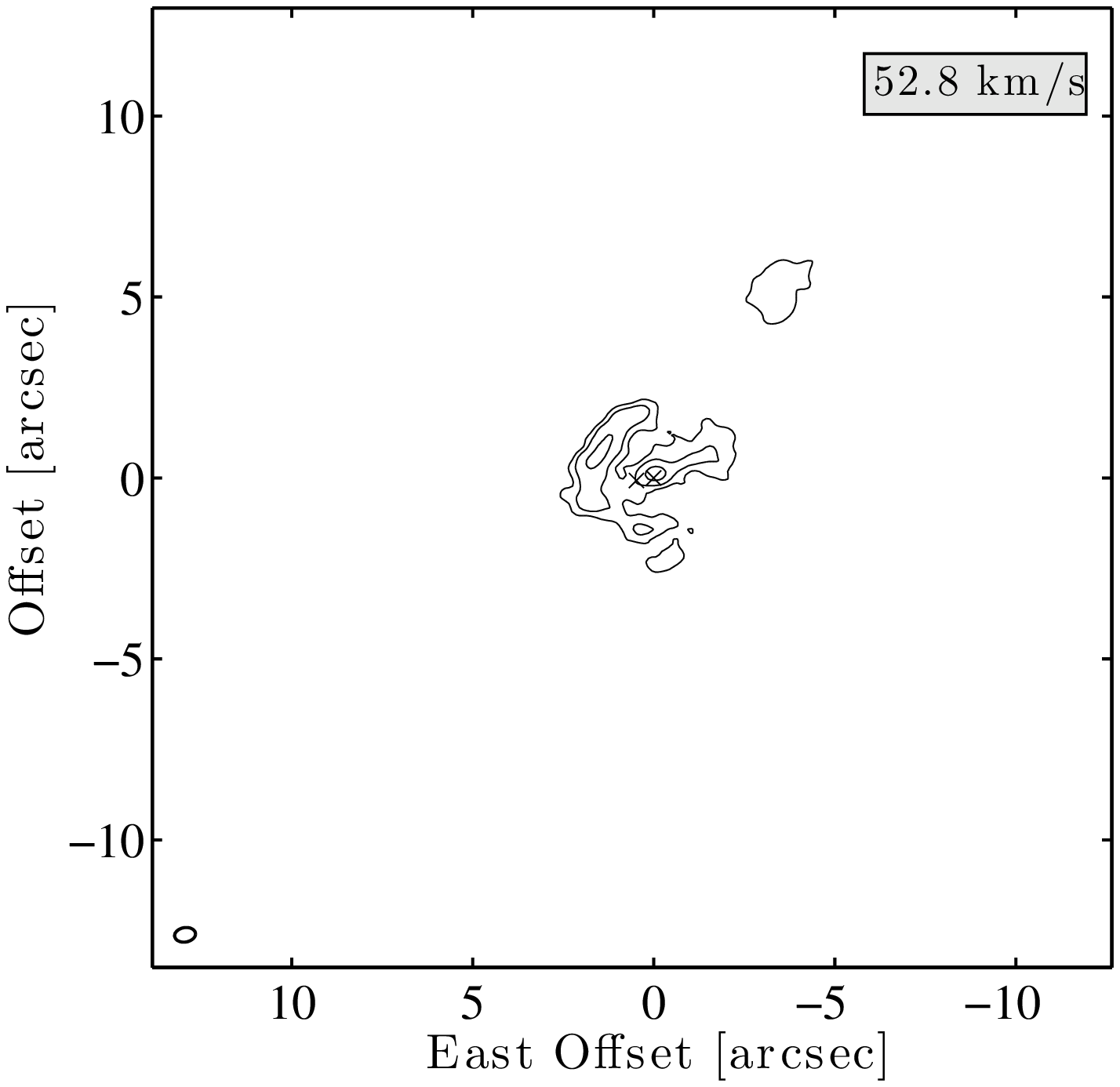} 
   \includegraphics[width=4.2cm]{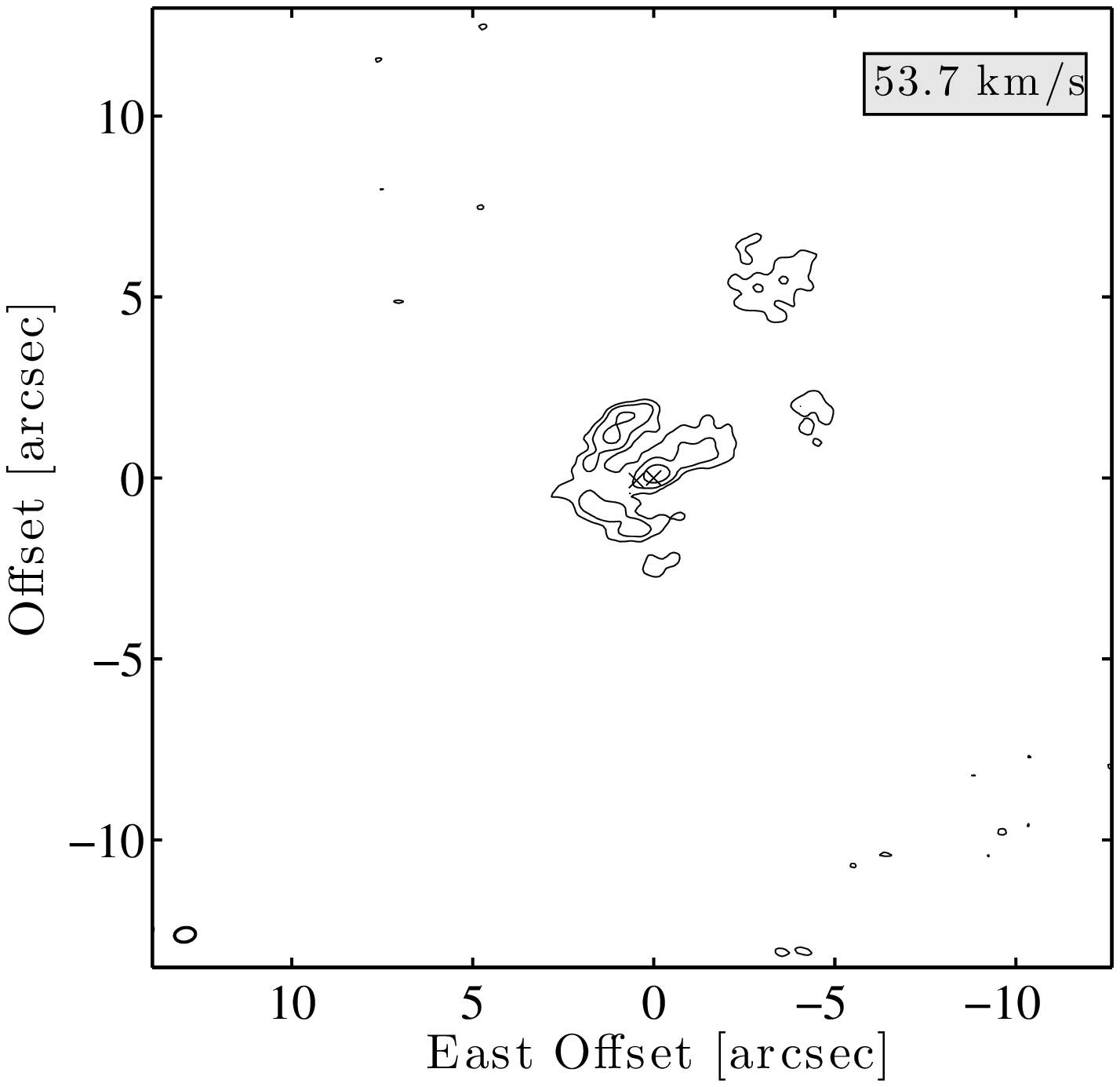} 
   \caption{The CO(3-2) map at 0.9\,km\,s$^{-1}$ resolution. The crosses mark the position of Mira A (west) and B (east). Contours are drawn at 3, 5, 10, 20, 40, and 80 times the rms noise level of the respective channel ranging from 30 to 160\,mJy. The beam is shown in the lower left corner.}
   \label{map}
\end{figure*}

\begin{figure}[htbp]
   \centering
   \includegraphics[width=\columnwidth]{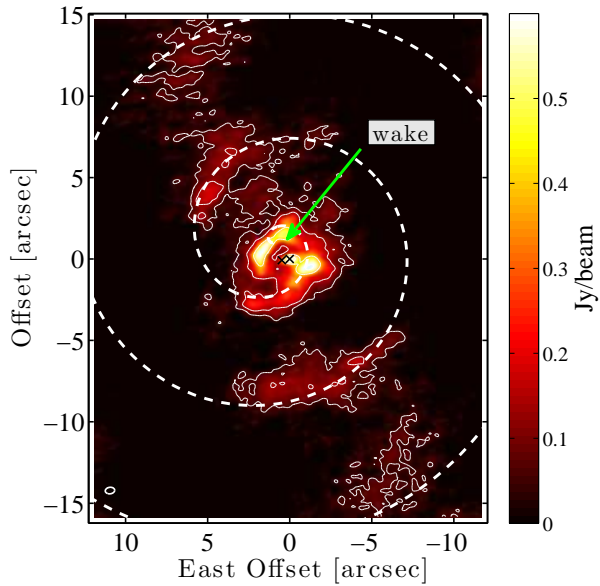} 
   \caption{The CO(3-2) map averaged from 49 to 54\,km\,s$^{-1}$. The crosses mark the position of Mira A (west) and B (east). Contours are drawn at 3, 5, 10, 20, 40, and 80 times the rms noise level of the respective channel ranging from 30 to 160\,mJy. The beam is shown in the lower left corner. A spiral is overplotted to clearly mark the spiral windings and the accretion wake behind the companion is indicated. }
   \label{spiralwake}
\end{figure}

Figure~\ref{spiralwake} shows an average from 49 to 54\,km\,s$^{-1}$ where the spiral windings and the accretion wake discussed in Sects~\ref{morph} and \ref{shape} are clearly marked. 

\end{document}